\documentclass{aastex6}
\fullcollaborationName{The Friends of AASTeX Collaboration}
\begin{document}

\title{Small Jupiter Trojans Survey with Subaru/Hyper Suprime-Cam}

\author{
Fumi Yoshida\altaffilmark{1,2}, 
Tsuyoshi Terai\altaffilmark{3}, 
}
\email{fumi.yoshida.ermei@gmail.com}
\footnote{Based on data collected at Subaru Telescope, which is operated by the National 
Astronomical Observatory of Japan.}

\altaffiltext{1}{Planetary Exploration Research Center, Chiba Institute of Technology, 
2-17-1 Tsudanuma,Narashino, Chiba 275-0016, JAPAN}
\altaffiltext{2}{Department of Planetology, Graduate School of Science, Kobe University, Kobe, 657-8501, Japan}
\altaffiltext{3}{Subaru Telescope, National Astronomical Observatory of Japan, National Institutes 
of Natural Sciences (NINS), 650 North A`ohoku Place, Hilo, HI 96720, USA}

%%%%%%%%%%%%%%%%%%%%%%%%%%%%%%%%%%%%%%%%%%%%%%%%%%%%%%%%%%%%%%%%%%%%%%%%%%%%%%%%%%%%%%%%%%%%%%%%%%%
\begin{abstract}

We observed the L4 Jupiter Trojans (L4 JTs) swarm using the Hyper Suprime-Cam attached to the 8.2~m 
Subaru telescope on March 30, 2015 (UT). 
The survey covered $\sim$26~deg$^{2}$ of sky area near the opposition and around the ecliptic plane 
with the 240-sec exposure time in the $r$-band filter through the entire survey. 
We detected 631 L4 JTs in the survey field with the detection limit of $m_r$~=~24.4~mag. 
We selected 481 objects with absolute magnitude $H_r$~$<$~17.4 mag and heliocentric distance 
$r$~$<$~5.5~au as an unbiased sample and then used them to estimate the size distribution. 
Assuming a geometric albedo of 0.07 \citep{grav12}, the size range of our unbiased sample is 
$\sim$~2~--~20~km in diameter ($D$). 
We can fit a single-slope power law to the cumulative size distribution and then found the best-fit 
index ($b$) is $b$~=~1.84~$\pm$~0.05 in $N(>D)~\propto~D^{-b}$. 
The slope value ($\alpha $) of corresponding absolute magnitude distribution 
($N(H)~\propto~10^{\alpha H}$) is 0.37~$\pm$~0.01. 
This $\alpha$ is consistent with that of the faint end slope presented by \cite{WB15}. 
%However, the entire size distribution is slightly different from the previous surveys performed by 
%\citet{YN05,YN08} and \citet{WB15} which reported a broken power-law size distribution of L4 JTs in 
%a similar size range, but with a small number of samples. 
%So far, this work provides the deepest wide-field survey for faint JTs. 
The size distribution obtained from this survey is slightly different from previous survey's result with a similar size range \citep*{YN05,YN08,WB15}, which reported a broken power-law or double power-law slopes in their cumulative size distribution.
Our results insists that the slope of $b$~=~1.84 continues from $H$~=~14.0 to at least $H$~=~17.4. 
Since this work contains the largest L4 JT samples and one magnitude deeper than the study by \cite{WB15}, we believe that our study obtained  the robustest size distribution of small JTs so far. 
Combining the cataloged L4 JTs and our survey, we finally show the entire size distribution of L4 
JTs up to $H_r$~=~17.4~mag. 

\end{abstract}
%%%%%%%%%%%%%%%%%%%%%%%%%%%%%%%%%%%%%%%%%%%%%%%%%%%%%%%%%%%%%%%%%%%%%%%%%%%%%%%%%%%%%%%%%%%%%%%%%%%

\keywords{minor planets, asteroids: general}



%%%%%%%%%%%%%%%%%%%%%%%%%%%%%%%%%%%%%%%%%%%%%%%%%%%%%%%%%%%%%%%%%%%%%%%%%%%%%%%%%%%%%%%%%%%%%%%%%%%

\section{Introduction} \label{sec:intro}

%\textcolor{red} 
The Small Solar System Bodies (SSSBs) are survivors of planetesimals which have not accumulated into the planets. 
Since most of planetesimals have not experienced significant thermal evolution so far, their chemical compositions can reflect initial materials of the proto-planetary disk of our solar system and their physical properties (size, shape, rotation period, etc.) have recorded probably their collisional evolution which have experienced so far. 
Moreover, their orbital distributions contain the results of gravitational interactions and dynamics evolution which they have experienced since they were formed.
Therefore, studying the SSSBs from multiple perspectives is indispensable and it is an unique way to fully understand the entire formation history of our solar system.

%}
%The Small Solar System Bodies (SSSBs) preserve an important record of gravitational interactions and dynamical evolution of the solar system in their orbits. 
%Their compositions and physical properties provide us clues for understanding the initial conditions of the early solar nebula. 
%Studying the SSSBs from multiple perspective is an essential and unique way to fully understand the history of our solar system. 

In the current solar system, the SSSBs are located separately in several groups. 
Among such groups, main belt asteroids (MBAs), trans-Neptunian objects (TNOs) and Jupiter Trojans 
(JTs) are the most populated groups. 
In this paper,
we focus on the JTs, because JTs are located in the middle region between most populated regions: the main asteroid belt and Kuiper belt, and they are supposed to be very important objects in connecting between inner and outer objects of the solar system.

JTs share their orbit with Jupiter.
They are located around the triangular stationary points, namely the Lagrangian points L4 (leading) 
and L5 (trailing) of Jupiter. 
Such special orbits and locations of JTs attracted people to study JTs as problems of 
mathematics or celestial mechanics at the time JT was first discovered. 
However, recently our interest in JTs has extended to their origin, physical properties, chemical 
composition and so on. 

%Since JTs are a significant population that lies between the main asteroid belt and Kuiper belt, they play a crucial role in interconnecting the inner and outer solar system bodies. 
%Therefore, the origin and evolution of the JT population are important keys for understanding the dynamical history of the solar system.

%Classical solar system formation models \citep{peale93, MS98a, MS98b} suggest that JTs have formed from the planetesimals in the vicinity of Jupiter around 5 au from the Sun. 
%They were captured into Trojan's orbits by Jupiter probably with gas drag of the solar nebula during Jupiter's mass growth. %However, newer \textcolor{red}{theoretical models of planet migration} (such as the ``Nice model" \citep{mor05} or ``Jumping  Jupiter model" \citep{nes13, RN15} suggest that JTs are the scattered objects from the outer region of the solar system as a result of gravitational interaction including resonant dynamics during the planet migration process that took place in the early solar system. If the newer models are right, JTs and scattered TNOs could have originated in the same region of the solar system.

In the 21th century JTs' importance in the solar system science became even larger, partly due to the rise of new theoretical models of planet formation, such as Grand Tack \citep*{wal11}, Nice \citep*{mor05}, or Jumping Jupiter \citep*{nes13,RN15}. 
These models collectively claim that planets, protoplanets, and planetesimals once radially migrated in the early solar system through their gravitational interaction  including resonant dynamics in the early solar system, causing a substantial radial stirring of material.
If such a mixing process really happened in the solar system, we must be able to find some evidence remaining in the current SSSBs. 
JTs are highly promising candidates of these objects that encompass evidence of radial stirring in the early solar system.
In this regard, understanding the origin and evolution of JTs is practically equivalent to understanding the dynamical history of the solar system in its early stage.

 %\textcolor{red}{
 \citet{FB12} found that distant object groups (Centaurs, scattered TNOs, Resonant TNOs, and Classical TNOs) have two different components in each group, which are red color objects and neutral color objects. The neutral color objects are common at all the groups, while the red colors components are different between groups.
They inferred that the neutral objects formed with the material broadly distributed in the outer protoplanetary disk and the red objects formed with different components depending on the heliocentric distance have existed before the violent scattering event happened in the primordial disk.   
%}
%This guess tells us that both groups would show similar physical properties (e.g., similar size distribution) and similar chemical composition (e.g., similar colors). 
%Therefore, comparative studies on size distributions and colors distributions between JTs and other SSSB groups can be a unique method and the best way to investigate their origins as well as what dynamical disturbance was like in the early solar system.
%The size distribution can be a fingerprint of dynamical environment of the independent population, and the color of SSSBs would be an indicator of where they were formed. 
%Therefore, we would be able to investigate the origin of JTs using size distribution and colors.
Recent observations revealed that there are two different groups in the JT population: the red 
group (R$_{\rm g}$) and the less red group (LR$_{\rm g}$) \citep*{emery11,wong14,WB15}, (hereafter WB2015). 
% \textcolor{red}{
A correlation between the two different color groups in the JTs population and the two components in each group of distant objects has not been revealed so far.
If the objects formed in the distant region have been implanted into the JT region as the planet migration models predicted, the surface color is likely to be altered by the temperature difference and/or different solar radiation environment (bring space weathering). This implies that the current surface color of objects cannot be no longer used as an indicator of the formation region of the objects. 
%This implies that the current surface color of objects no more can be used as a indicator of the formation region of the objects.

%%%%%%%%%%%%%%
Therefore we focus on the size distribution of the SSSB group. 
A scattering process by planet migration is independent to the size of objects scattered by close approach with the planets. If a planet scattered a group of SSSBs and then the scattered objects made a new group of SSSBs, the original size distribution of previous group would be copied to that of new group. 
Therefore, we can say that the size distribution is insensitive to dynamical environment and dynamical disturbance, rather it is mainly determined by a condition of the formation environment of the objects, and is altered by subsequent collisional evolution in the object group \citep{davi02, OBG05}.
 \citet{bott05} simulated a collisional evolution of the MBAs and then suggested that the size distribution of MBAs was quickly evolved and the wavy-shape of the size distribution of MBAs is a fossil from the violent early epoch, namely the accretion phase of proto-planets. 
\citet{strom05} found that the size distribution of old Lunar craters, which were formed 3.8 Gyr ago when the late heavy bombardment (LHB) happened, closely resembles the size distribution of current inner MBAs. This means that the size distribution of MBAs had reached to its present shape 3.8 Gyr ago (equal to the timing of LHB), suggesting that the collisional evolution in the main belt in the last 3.8 Gyr is not so effective to change the shape of size distribution.
Similarly, if the collisional evolution haven't changed the shape of size distribution significantly of the JTs and other SSSB groups in the last 3.8 Gyr, comparative study of the size distribution between JTs and other SSSBs is a strong tool to confirm their identical origin.
A main cause to determine the entire shape of the size distribution of each SSSB group is a strength law which reflects the inner structure or the composition of objects evolved by collisional evolution.
Therefore, if the size distribution is similar between the SSSB groups, the groups have a similar strength law, namely their inner structure or the composition is similar, we can say that they likely have the same origin.

In this paper, we show the size distribution of small L4 JTs obtained from Subaru/Hyper Suprime-Cam (HSC) \citep{miya12, miya13}, and then we compared our result with \cite{YN05} and WB2015 that had performed a similar observation of L4 JTs using Subaru/Suprime-Cam \citep{miya02}. 
We describe our observation in section~2 and our data analysis in section~3. 
The newly obtained size distribution of L4 JTs is illustrated in section~4. 
In the final section, section 5, by comparing the size distribution between different dynamical groups in the current solar system, we briefly discuss a  dynamical history of the solar system.

%%%%%%%%%%%%%%%%%%%%%%%%%%%%%%%%%%%%%%%%%%%%%%%%%%%%%%%%%%%%%%%%%%%%%%%%%%%%%%%%%%%%%%%%%%%%%%%%%%%
\section{Observations} \label{sec:obs}

The survey observations of the L4 JTs were performed by the 8.2-m Subaru Telescope which was 
equipped with HSC at the prime focus on March 30, 2015 (UT). 
HSC is a gigantic mosaic camera containing 116 2k~$\times$~4k Hamamatsu fully depleted CCDs 
(104 for science, four for autoguiding, and eight for focus monitoring) with a field-of-view (FOV) 
of 1.5~degrees in diameter,  which is about six times larger than the FOV of the Suprime-Cam 
\citep{miya02}, and a pixel scale of 0$\farcs$17 \citep{miya12}.

The survey area consists of 17~FOVs of HSC, corresponding to $\sim$26~deg$^{2}$ of sky, centered 
at RA~=~12$^{\rm h}$33$^{\rm m}$ and Dec~=~$-$03$\arcdeg$00$\arcmin$ (see Figure~\ref{fig01}). 
The field was situated within 5$\arcdeg$ from the opposition and within 4$\arcdeg$ from the 
ecliptic plane, as well as 10~--~20$\arcdeg$ from the L4 point of Jupiter. 
%\textbf{
%The slight difference in ecliptic longitude from the opposition is the essential condition for estimating the orbits of detected JTs with less uncertainty. 
%}
% \textcolor{red}{
In order to estimate asteroid's orbit from the apparent motion of them with a short observational arc, it is necessary to measure the asteroid's velocity at/near the opposition. 
On the other hand, in order to find many JTs at once, it is better to observe near the L4 point as expected from a spatial distribution of known JTs. Therefore, we hoped to perform our survey when the L4 point close to the opposition. 
However, the Subaru telescope schedule did not allow us to do so, we had to choose either area near the opposition or near the L4 point. Since we have to distinguish JTs from Hildas or outer MBAs without contamination for obtaining the size distribution of each dynamical group separately, we preferred to survey near the opposition rather than near the L4 point. We finally selected our survey area near the opposition avoiding very bright stars, where was a little shifted (10~--~20$\arcdeg$) from the L4 point.
%}
All of survey area are covered by the Sloan Digital Sky Survey (SDSS) field \citep{alam15} for 
astrometric/photometric calibrations.

All of the data were obtained in the $r$-band with an exposure time of 240~sec. 
Each field was visited three times with a time interval of less than one hour, as shown in 
Table~\ref{tab01}. 
The average seeing size of each field was 0$\farcs$6 to 1$\farcs$0, but larger than 1$\farcs$2 at 
the two fields, FIELD10 and FIELD11.

%%%%%%%%%%%%%%%%%%%%%%%%%%%%%%%%%%%%%%%%%%%%%%%%%%%%%%%%%%%%%%%%%%%%%%%%%%%%%%%%%%%%%%%%%%%%%%%%%%%
\section{Data Analysis} \label{sec:ana}

\subsection{Image reduction} \label{sec:ana1}

The data are processed with the HSC data reduction/analysis pipeline, named \textit{hscPipe} 
(version~3.8.5), developed by the HSC collaboration team based on the Large Synoptic Survey 
Telescope (LSST) pipeline software \citep{ivez08, axel10}. 
We run only the single-frame processing, i.e., the data processes for each exposure including 
image correction, source detection, measurement, and calibration.
First, the images are reduced by standard procedures such as bias subtraction, trimming of the 
overscan/prescan regions, flat-fielding, defect removal, and sky background subtraction. 
%\textbf{
Next, the pipeline detects sources from the corrected images, performs source measurements 
of centroids, shapes, and photometry with several methods, and compares the coordinates and fluxes 
with the matched objects in a reference catalog for astrometric and photometric calibration. 
%}
We use the SDSS DR9 catalog with the source fluxes corrected by the data of Panoramic Survey 
Telescope and Rapid Response System (Pan-STARRS)~1 (PS1) survey \citep{schl12, tonr12, magn13}.
%\textbf{
The PS1 $r$ magnitude ($r_{\rm PS1}$) is converted into the HSC $r$ magnitude ($r_{\rm HSC}$) 
using a color term equation determined by HSC software team as 
\begin{equation}
r_\mathrm{HSC} = r_\mathrm{PS1} + 0.00280 + 0.02094\, ( r_\mathrm{PS1} - i_\mathrm{PS1} ) 
- 0.01878\, ( r_\mathrm{PS1} - i_\mathrm{PS1} )^2 . 
\end{equation} 
%}
Finally, \textit{hscPipe} estimates the astrometric solution and photometric zero-point
(in the AB magnitude system), then creates a source catalog with the measured values (e.g., 
centroids, shapes, fluxes) for each CCD image.

\subsection{Detection} \label{sec:ana2}

We used the source catalogs produced by \textit{hscPipe} for mechanical detection of moving objects. 
We developed an algorithm allowing us to efficiently extract moving objects from the source 
catalogs with the following procedures: 
(i)~Removing suspected cosmic rays and saturated sources based on the flags added by the 
pipeline processes; 
(i\hspace{-.1em}i)~Excluding the sources which have identical coordinates among three visits as 
stationary objects; 
(i\hspace{-.1em}i\hspace{-.1em}i)~Searching for combinations of sources from each visit 
whose positions have a pattern corresponding to uniform linear motion on the sky. 

For selective capture of JTs and Hildas, we limit the motion range to be searched covering 
their motion distribution derived from orbits with eccentricity of 0.0--0.3 and 
inclination of 0$\degr$--40$\degr$ at the survey field, as shown by the area enclosed with the 
dotted black lines in Figure~\ref{fig02}.
We regard the detected moving objects with motion vectors located on this area as JT/Hilda 
candidates.
Finally, all of the image of candidate sources are visually inspected. 

The apparent velocities along ecliptic longitude/latitude of the detected JT/Hilda candidates are 
plotted in Figure~\ref{fig02}. 
We overplotted motion distributions of synthetic orbits for JTs and Hildas randomly generated 
from probability distributions based on the orbital distributions of known objects obtained from 
the Minor Planet Center (MPC) database\footnote{\url{http://minorplanetcenter.net/db_search}} in  Figure~\ref{fig02}.
One can see that the detected objects divide into two swarms and the motion distribution well 
matches that of the synthetic JT/Hilda objects. 
We defined the boundary between these two groups as motions of circular orbits with semi-major 
axis of 4.5~au (the broken red line in Figure~\ref{fig02}). 
The slower group corresponding to JTs consists of 631 objects, while the faster group corresponding 
to Hildas consists of 130 objects. 

In this paper, we focus on the investigation of JTs' size distribution. 
The analysis for the Hilda candidates will be reported in another paper (Terai \& Yoshida in 
preparation).

\subsection{Measurements} \label{sec:ana3}

As mentioned in the previous section, \textit{hscPipe} measures various parameters of the
detected sources including centroid positions. 
However, the centroid for a moving object with an elongated shape may be inaccurate because the 
barycenter of intensity profile is sensitive to instability of seeing and/or transparency during 
the exposure. 
Therefore, instead of using the \textit{hscPipe} measurement, we determine independently the center positions of the detected JTs by making a $\chi^2$ fitting of object models to the image data.
The model is generated from integration of Gaussian profiles with the center shifting with a constant motion equal to the measured velocity, as shown in Figure~\ref{fig03}. 
The FWHM of the individual Gaussian profile is given as the typical seeing size of the image. 

Using the fixed center position, we measure the total flux of each object by aperture photometry 
with the sinc-interpolation technique developed by \citet{BL13}. 
The aperture flux is computed from a weighted sum of pixel values over an aperture $A$ as
\begin{equation} 
 f_A = \sum_{i,j} \ w_{ij} \ p_{ij}, 
\end{equation} 
where $p_{ij}$ is the discretely sampled flux given from a continuous flux function $p(x, y)$. 
$w_{ij}$ is the weighting term represented by
\begin{equation} 
 w_{ij} = \int \!\!\! \int_A \ \frac{\sin(\pi(x-x_i))}{\pi(x-x_i)} 
 \ \frac{\sin(\pi(y-y_j))}{\pi(y-y_j)} \ dxdy. 
\end{equation} 

The aperture is formed into a shape of trailed image with a radius of 2.0$\arcsec$ with the velocity of each moving object 
%elongated by the measured motion 
(see Figure~\ref{fig03}).
If there is a star/galaxy within the aperture area, we perform the same photometry again after 
background subtraction with a reference image of another visit via point-spread-function matching. 
The measured flux is converted into apparent magnitude using the photometric zero-point estimated 
by \textit{hscPipe}.
The representative magnitude is determined as the weighted mean of the values at all the visits
with the photometric errors.

%\textbf{
 As shown in Table~\ref{tab01}, the time intervals of exposures are $\sim$18~min and 36--56~min
between the first and second, and the second and third visits, respectively.
The brightness variation due to asteroid rotation could cause an additional uncertainty in the
apparent magnitude measurement.
To assess the effect of asteroid rotation,  we estimated the magnitude differences among
the visits, $\Delta m_r = m_{r,i+1} - m_{r,i} \; (i = 1, 2)$ where $m_{r,i}$ is the $r$-band
magnitude at the $i$-th visit, for each object as an index representing the both photometric and
rotational contributions \citep{WB15}.
As performed in WB2015, the standard deviations of $\Delta m_r$ contained in 0.5~mag bins
of absolute magnitude (see Section~\ref{sec:res-1}) were compared with the medians of photometric
errors of the same objects.
We found that there is no significant difference between the two values in the most bins,
indicating a rotational contribution of $\sim$0.05~mag or less.
Since this is much smaller than the uncertainty of the estimated heliocentric distance
corresponding to 0.08~mag for absolute magnitude (see Section~\ref{sec:ana4}), the size
distribution is likely not to be affected by the rotational contribution.
Thus, we exclude the effect of rotational magnitude variation from consideration in this survey.
%}

\subsection{Orbits} \label{sec:ana4}

The observation arcs of the detected JTs are less than 80~min, too short to determine their 
orbits. 
However, the survey field was located close to the opposition, allowing us to approximate the 
orbital elements from the measured sky motions assuming circular orbits, with relatively 
small uncertainties. 
We estimated the semi-major axis (or heliocentric distance) and inclination of each object 
using the expressions presented in \citet{terai13}. 

It is necessary to evaluate the accuracy of heliocentric distance in our orbit approximation 
since the value is used for calculation of the absolute magnitude. 
We conducted a Monte Carlo simulation of the orbit fitting using the synthetic object generator 
described in Section~\ref{sec:ana2}. 
Figure~\ref{fig04}a shows a comparison of heliocentric distance between the generated and 
estimated orbits. 
It displays that the estimated heliocentric distances have a systematic deviation from the 1:1 
line. 
This is because the sky motion of an elliptically orbiting object at the perihelion side is slower 
than that of a circularly orbiting object with the equal heliocentric distance, while the sky 
motion of an elliptically orbiting object at the aphelion side is faster than that of a circularly 
orbiting object with the equal heliocentric distance.
The correlation between the actual and estimated heliocentric distances ($r_{\rm act}$ and 
$r_{\rm est}$, respectively, in au) is fitted with a linear function as 
$r_{\rm est} \, = \, 0.521 \, r_{\rm act} \, + \, 2.48$ (a solid line in Figure~\ref{fig04}a). 
Figure~\ref{fig04}b is the same plot as Figure~\ref{fig04}a but with the estimated heliocentric 
distances corrected by this function. 
The systematic error can be removed from the estimated heliocentric distances of the detected 
JTs by applying this correction. 
The statistical error of the corrected heliocentric distance is 0.09~au. 
This causes uncertainties in the absolute magnitude and body diameter of 0.08~mag and four 
percent, respectively.

\subsection{Detection efficiency}

For accurate investigation of the size distribution of the JT population, it is critical to evaluate the completeness of the  detection as function of apparent magnitude. 
The sensitivity depends on the vignetting effect increasing with a distance from the FOV center 
as well as the time variation of airmass and sky condition. 
We examined the detection efficiency for JTs in all the frames on a CCD-by-CCD basis with the 
following processes: 
(i)~Creating a synthetic blank image reproducing the sky background of the original image; 
(i\hspace{-.1em}i)~Implanting synthetic moving objects corresponding to JTs with a given flux 
produced in the same manner described in Section~\ref{sec:ana3} into the generated image at (i); 
(i\hspace{-.1em}i\hspace{-.1em}i)~Processing this image with the reverse order of the data 
reduction procedure as a pseudo raw image; 
(i\hspace{-.1em}v)~Running \textit{hscPipe} for the pseudo raw image and counting the number 
of detected JT sources. 

The above procedure is repeated for synthetic JTs with apparent magnitude $m_r$ from 24.0~mag 
to 25.4~mag in increments of 0.2~mag. 
We represent the detection efficiency curve for a single CCD image with a sum of the two functions 
for proper fitting as 
\begin{equation}
\eta (m) = \sum_{k=1,2} \frac{\epsilon_k (m)}{2} 
           \left[ 1 - \tanh \left( \frac{m - m_{50}}{w_k} \right) \right], 
\end{equation}
where $m_{50}$ is the magnitude at half maximum and $w_k$ is the transition width. 
$\epsilon_k (m)$ shows the fraction of each term, given by 
\begin{equation}
\epsilon_1 (m) = 1 - \epsilon_2 (m) = \frac{1}{2} 
                 \left[ 1 - \tanh \left( \frac{m - m_{50}}{w} \right) \right]. 
\end{equation}

Actually, the necessary conditions for identifying a moving object is to detect it from all of 
the three visits. 
The expected detection efficiency for JTs in each field is given by
\begin{equation}
\eta (m) = \prod^{N}_{i=1} \eta_i (m), 
\end{equation}
where $\eta_i (m)$ is the individual detection efficiency of the $i$-th visit ($N$~=~3). 

Figure~\ref{fig05} shows the detection efficiency of the CCDs located around the center,
middle, and the edge of HSC's FoVs (red, green, and blue lines, respectively)
taken at all of the 17 survey fields.
You can see the lowest detection efficiency occurred at the one of CCD near
the edge.
We defined the lowest detection efficiency of 0.5 (i.e. 50\% detection) as a
detection limit of our survey, which is 24.4~mag in apparent magnitude.

%%%%%%%%%%%%%%%%%%%%%%%%%%%%%%%%%%%%%%%%%%%%%%%%%%%%%%%%%%%%%%%%%%%%%%%%%%%%%%%%%%%%%%%%%%%%%%%%%%%
\section{Results} \label{sec:res}

\subsection{Sample selection} \label{sec:res-1}

%\textbf{
The absolute magnitude of the detected JTs in the $r$ band is estimated from the equation presented
by 
\citet{bowe89}:
\begin{equation}
H_r = m_r - 5 \log ( r \Delta ) - P(\theta),
%H_r = m_r - 5 \log ( r \Delta ) + 2.5 \ log [ (1-G) \Phi_1 + G \Phi_2 ],
\end{equation}
where $r$ and $\Delta$ are the heliocentric and geocentric distances, respectively. $G$ is the slope parameter, and  P($\theta$) is s the phase function at a phase angle $\theta$ given by
\begin{equation}
P(\theta) = -2.5 \log [ (1-G) \Phi_1 + G \Phi_2 ].
\end{equation}
$\Phi_{1}$ and $\Phi_{2}$ are is the phase function at a phase angle $\theta$ given by 
\begin{equation}
\Phi_i = \exp \left(-A_i [ \tan (\theta /2)]^{B_i} \right), i=1,2,
A_1=3.33, A_2=1.87, B_1=0.63, B_2=1.22. 
\end{equation} 
%}
In our survey, we could not measure the P($\theta$) of each object bexause our survey was done at only single phase angle. Therefore we assumed a constant slope parameter of $G$~=~0.15.
% as in the MPC catalog.

The absolute magnitude can be converted into body diameter $D$ by 
\begin{equation}
\log D = 0.2 m_{\sun,r} + \log ( 2 r_\earth ) - 0.5 \log p - 0.2 H_r, 
\end{equation}

where $m_{\sun,r}$ is the apparent $r$-band magnitude of the Sun \citep[$-$26.91~mag;][]{fuku11},
$r_\earth$ is the heliocentric distance of the Earth (i.e., 1~au) in the same unit as $D$, and 
$p$ is the geometric albedo. 
We assumed a constant albedo of $p$~=~0.07 for JTs based on an analysis of the NEOWISE data 
\citep{grav12}, which shows the mean albedo of 0.07~$\pm$~0.03 across all sizes, consistent with 
the C, P, and D taxonomic classes in JT population. 

Figure~\ref{fig06} shows the plot of the heliocentric distance and absolute magnitude.
The broken line represents apparent magnitude of the detection limit, $m_r$~=~24.4~mag.
For avoiding a detection bias caused by the decrease in brightness with increasing distance from 
the Sun and Earth, we defined the outer edge of JTs as $r$~=~5.5~au, where $m_r$~=~24.4~mag 
corresponds to $H_r$~=~17.4~mag, and selected objects located in the region of $r$~$\leq$~5.5~au 
and $H_r$~$\leq$~17.4~mag as an unbiased sample.
The extracted sample contains 481~objects.
%\textbf{
Compared with WB2015 who detected over 550~JTs but only analyzed an unbiased sample of 150 objects
with $H_V$~=~7.2--16.4~mag ($H_V$ is the absolute magnitude in $V$ band), our survey obtained more
than three times as many unbiased sample JTs as the one.
%}

\subsection{Size distribution} \label{sec:res-2}

Figure~\ref{fig07} shows the cumulative size distribution (CSD) of JTs in the unbiased sample as a 
function of $H$ magnitude with a bin width of 0.5~mag, which is sufficiently larger than the 
typical $H_r$ uncertainty ($\lesssim$~0.2~mag). 
The cumulative number is corrected by the detection efficiency as 
\begin{equation}
N ( < H ) = \sum_{ j : H_j < H } \frac{ 1 }{ \eta (m_j) }, 
\end{equation}
where $m_j$ and $H_j$ are the apparent and absolute magnitudes of object~$j$, respectively. 
%\textbf{
The error bars are given by the Poisson statistics.
%}

We found that the CSD in $H_r$~$>$~13.0~mag can be represented by a single-slope power law, not a 
broken power law as claimed by WB2015. 
This allows the differential $H$ distribution, $\Sigma ( H ) = dN(H)/dH$, to be fitted by
\begin{equation}
\Sigma ( H ) \, = \, 10^{\alpha ( H - H_0 )}, 
\label{eq_sgm-h}
\end{equation}
where $\alpha$ is the power-law slope, and $H_0$ is given as $ \Sigma ( H_0 ) = 1$. 
When the CSD is expressed as 
\begin{equation}
N ( > D ) \propto D^{-b}, 
\end{equation}
where $N(>D)$ is the number of objects larger than $D$ in diameter, the power-law index $b$ is 
converted into $\alpha$ by $b = 5 \alpha$. 

We used a maximum likelihood method \citep[e.g.,][]{bern04} for fitting Equation~(\ref{eq_sgm-h}) 
to the $H$ distribution of the unbiased JT sample. 
The likelihood function is given by 
\begin{equation}
L ( \alpha, H_0 ) \, \propto \, e^{-\tilde{N}} \prod_j \, \eta ( m_j ) 
                             \, \Sigma ( H_j | \alpha, H_0 ).
\end{equation}
$\tilde{N}$ is the expected number of detected objects in this survey, which is estimated by 
\begin{equation}
\tilde{N} = \int \, dH \, \eta ( m ( H ) ) \, \Sigma ( H | \alpha, H_0 ), 
\end{equation}
where $m(H)$ is the apparent magnitude approximated as $H$~+~5~$\log (r \Delta)$ with 
$r$~=~5.2~au and $\Delta$~=~4.2~au. 
Uncertainties in the fitted parameters is estimated from repeated fitting to synthetic object
samples generated from the actual objects based on the measurement errors. 

We obtained the best-fit power-law slope of $\alpha$~=~0.37~$\pm$~0.01, corresponding to 
$b$~=~1.84$~\pm~$0.05. 
This value agrees with the result of \citet{YN05}, $b$~=~1.89~$\pm$~0.10 in $D$~=~2--10~km, 
as well as the faint-end slope with $\alpha$~=~0.36$^{+0.05}_{-0.09}$ in 
$H_V$~=~14.9--16.4~mag shown by WB2015, though the precision is highly improved compared 
with those previous studies. 

The best-fit power law is plotted in Figure~\ref{fig07} as a broken line. 
The data is very close to the fitted line over $H_r$~$\gtrsim$~13.0~mag, indicating no evidence 
of the power-law break at $H_V$~=~14.93$^{+0.73}_{-0.88}$ reported by WB2015. 
We concluded that L4 JTs have a single-slope power-low size distribution in the range of
13.0~$\lesssim$~$H_r$~$\lesssim$~17.0. 

Finally, we compare the CSD obtained from our sample with that of known L4 JTs from the MPC 
catalog consisting of 4,080 objects. 
%\textbf{
Figure~\ref{fig08} shows that MPC JTs exhibit an almost constant increase with absolute magnitude
in log scale over $H_V$~$\sim$~10.0--14.0~mag, indicating that the completeness limit is likely to
be located at $H_V$~$\sim$~14.0~mag. 
We assumed the $V-r$ color of 0.25~mag \citep{szabo07} and scaled the CSD of our sample to the
cumulative number of the MPC sample at $H_V$~=~14.0~mag.
We also add the CSD from the L4 JT sample containing 93 objects presented by 
\citet[][hereafter JTL]{jew00}, which is normalized at $H_V$~=~15.0~mag.
%}

As seen in Figure~\ref{fig08}, the three CSDs match each other well except for the faint 
ends of the MPC and JTL samples that seem to reach beyond the completeness limit. 
%\textbf{
The combination of those CSDs shows that there is likely to be a power-law break around
$H_V$~=~13~mag. 
We fitted a broken power law to the combined CSD with $H_V$~$>$~12.0~mag.
The function is given as
\begin{equation}
\Sigma ( H ) \, = \left \{
\begin{array}{ll}
 10^{\alpha_1 ( H - H_0 )}, &\quad \mathrm{for} \;\; H < H_\mathrm{break}\\
 10^{\alpha_2 H + ( \alpha_1 - \alpha_2 ) H_\mathrm{break} - \alpha_1 H_0},
 &\quad \mathrm{for} \;\; H \geqslant H_\mathrm{break},
\end{array}
\right.
\end{equation}
where $\alpha_1$ and $\alpha_2$ are power-law slopes for the brighter and fainter objects,
respectively, with the break magnitude $H_{\rm break}$ as a border.
We found the best-fit parameters of $\alpha_1$~=~0.50~$\pm$~0.01, $\alpha_2$~=~0.37~$\pm$~0.01,
and $H_{\rm break}$~=~13.56$^{+0.04}_{-0.06}$ (see Figure~\ref{fig09}).
The goodness of fit is evaluated by the Anderson-Darling test \citep{AD52,pre92},
\begin{equation}
D = \max_{-\infty < x < \infty}
    \frac{ \left| S(x) - P(x) \right| }{ \sqrt{ P(x) \bigl[ 1 - P(x) \bigr] } },
\end{equation}
where $S(x)$ is the observed CSD and $P(x)$ is the fitted function.
The statistic $D$ was calculated to be 0.078, which cannot reject the null hypothesis for equality
between $S(x)$ and $P(x)$ even at 40\% significance level.
%}

%%%%%%%%%%%%%%%%%%%%%%%%%%%%%%%%%%%%%%%%%%%%%%%%%%%%%%%%%%%%%%%%%%%%%%%%%%%%%%%%%%%%%%%%%%%%%%%%%%%
\section{Discussion} \label{sec:dis}
\subsection{Previous surveys}
The property of size distribution of small solar system bodies has been evaluated by the index of power-law distribution 
($\alpha$ or $b$, in equations (10, 11)). 
The first survey for the size distribution of L4 JTs was conducted by JTL using the Univ. of Hawaii 2.2 m telescope. 
This survey determined the index, $b$~=~2.0, or $\alpha$~=~0.4 in the range of 
$H$~=~11--16~mag. At their survey period, the number of cataloged JTs was still small. 
The ASTORB catalogue was only completed up to $H_V$~=~9 or so. 
Therefore, there was a gap in the size range of size distributions between the cataloged JTs and 
small JTs detected from the survey by JTL.
Later, \citet{szabo07} investigated the SDSS data (the third release of Moving Object Catalog; 
MOC3) and found that the index, $b$~=~2.2, or $\alpha$~=~0.44 in the range of $H_V$~=~10--13.5~mag 
and they determined that all JTs with $H_V$~$<$~12.3~mag had already been discovered and listed in 
the ASTORB file. 
By combining the ASTORB file, SDSS/MOC3 and the survey result by JTL, the size 
distribution of L4 JTs had been revealed up to 
%\textcolor{red}
{$H_V \sim$~16~mag.
%} 
Up to here, the surveys for JTs were done by 2 m class telescopes. 

%\textcolor{red}{
In order to examine the size distribution of JTs ranging smaller size than JTL, \citet{YN05} searched for small JTs in the data set  taken by Subaru telescope in February 2001. This is equivalent to the first L4 JTs survey done by 8 m class telescope \citep{yoshi03}.
%In February 2001, \citet{YN05} conducted the first L4 JT survey for detecting smaller JTs by the 8.2m Subaru telescope}. 
%The detected \textcolor{red}{JT's} range was 14~$<$~$H_V$~(mag)~$<$~17.7, corresponding to 2~$<$~$D$~(km)~$<$~10 (assuming an albedo of 0.04). 
They detected JTs ranging 14~$<$~$H_V$~(mag)~$<$~17.7, corresponding to 2~$<$~$D$~(km)~$<$~10 (assuming an albedo of 0.04) and estimated the cumulative size distribution.
The slope index $b$ was 1.89 for the entire size range. 
They also suggested that the size distribution may have a break around $H_V$~=~16~mag. In this case, the 
slope index changes from $b$~=~2.39 for $H_V$~$<$~16 mag to $b$~=~1.28 for $H_V$~$>$~16~mag.
%\textcolor{red}
%{\citet{YN08}  did a second survey \textcolor{red}{by the Subaru telescope} for finding the L5 JTs in October, 2001.}

\citet{YN08} used a data set of MBA survey taken by Subaru telescope in October 2001 for searching L5 JTs \citep{YN07}. This is equivalent to the second survey of JTs by Subaru telescope, but the first survey for L5 JTs.
Then they noticed that the L4 and L5 swarms may have different size distributions. 
%\textcolor{red}{
Although both surveys were important to determine the faint end of the size distribution of L4 and L5 JTs, the number of detection of JTs in both surveys were small : 51 L4 Trojans and 62 L5 Trojans. This is because the surveys were not dedicated survey for JTs, but the main purpose was for detection of MBAs. Therefore, we guess that their determination accuracy of the size distribution cannot be so high, because of small number statistics. A robust results by using large sample of JTs have been long-awaited.
%}

Recently, WB2015 conducted a dedicate survey for L4 JTs by Subaru/Suprime-Cam.
JTs ranging of 7.2~$<$~$H_V$~(mag)~$<$~16.4 were detected in their survey. 
By using the cataloged L4 JTs and their detected L4 JTs, they found that there are two break 
points in the cumulative size distribution of JTs up to $H_V$~=~16.4~mag : the first break at $H_{b1}$~=~8.46 and the second break at
$H_{b2}$~=~14.9, and the power-law slopes of $bs$ are 4.55$^{+0.95}_{-0.80}$ for the largest 
populations of L4 JTs, $\sim$~2.2 for the middle size objects, and 1.80$^{+0.25}_{-0.45}$
for the faint end of L4 JTs population (see Figure~\ref{fig07} in WB2015).
%%%%%
In this work, meanwhile, the break point in the cumulative distribution correspond to the second break of $H_{b2}$ in WB2015 was slightly shifted to a brighter magnitude ($H_{break}$=13.56) and the slope index at the faint end was $b$=1.84$\pm$0.05. The slope indexes between WB2015 and this work are consistent each other within the range of error bars. 
The break point at $H$=16 found by \citet{YN05} was not seen in WB2015 and this work. It seems to be a fluctuation induced from small number statistics.
%%%%%

%
\citet{wong14} who investigated the SDSS data set found that the R$_{\rm g}$ and the LR$_{\rm g}$ in the JT swarm have different size distributions, and WB2015 found most of smaller JTs in their sample belong to the LR$_{\rm g}$. Based on these findings, WB2015 proposed a possibility that the fragmentation of the red objects by collisional evolution creates the less-red objects and appeared into the difference of the size distributions. However, there is another possibility that the difference of the size distribution in the different color groups caused by the difference of chemical materials related to a difference of their formation region. 
For example, \citet{YN07} have reported that the S- and C-complex with the faint MBAs ($H_R$~$\gtrsim$~15~mag) detected by Subaru telescope with the Suprime-Cam  show the different size distributions. This finding probably reflects a difference in compositions or origins of the two groups in the population.

We could not confirm WB2015's findings, because our HSC survey used only the $r$-band for detecting JTs. This is because we wanted to find  JTs as many as possible during one night observation (the filter exchange of the HSC takes about 40 min). Further multicolor observations are needed to understand collisional evolution in a population which consisted of different taxonomic types or different composition groups. 

\subsection{Comparative studies of size distribution of JTs with other populations}
Since we estimated the size distribution for the smaller JTs by using the small 481 JTs with $H_R$~$\gtrsim$~17.4 mag, corresponding to $D~\lesssim$~2 km (assuming the albedos of 0.07 \citep{grav12}) detected by Subaru + Hyper Suprime-Cam, now, our knowledge on the size distribution of JTs has reached down to $D$~$\sim$~2 km. Therefore, we can compare the size distribution between JTs and MBAs down to 2 km in diameter. 

%\textcolor{red}{
For more detail comparison between the size distributions of JTs and MBAs, we made R-plots (Relative plot). This method was devised by the Crater Analysis techniques Working Group \citep{arvi79} to better show the size distribution of craters. When sufficient data set are available, the R-plot provides a more sensitive comparison between size distributions than the cumulative plot.  \cite{strom05, strom15} compared the size distributions between near Earth asteroids (NEAs), MBAs and Moon's young/old craters on the R-plots and then they found that impactor population which made Lunar highland craters in old era are different from the population which made Lunar mare craters in relatively young era and then they concluded that the source of impactors in the inner solar system region had been changed from MBAs to NEAs around 3.8 Gyr ago. Here we used the same method using in \cite{strom05, strom15} and made the R-plots for  Inner (2.0~$<$~$a$ (au)~$<$~2.6), Middle (2.6~$<$~$a$ (au)~$<$~3.0), Outer (3.0~$<$~$a$ (au)~$<$~3.5) MBAs and JTs separately in Figure~\ref{fig10}. We used the same data sets of MBAs with \cite{strom05, strom15} and, in addition to those, we added the data sets newly obtained by Subaru telescope  \citep{yoshi11}, AKARI survey \citep{usui11} and WISE survey \citep{masi11} to the R-plots. Please see the caption of Figure~\ref{fig10} for details for used dataset. In Figure~\ref{fig10}, the scale of vertical axis is arbitrary. We connected all data sets smoothly. 
%%%%%%%%20170506%%%%%%%%%%%%
At first, we noticed that a wavy structure seen in the R-plot of MBAs does not show up on the R-plot of JTs. 
According to the numerical simulations of collisional evolution
for L4 JTs by \cite{deEB07}, no-wavy structure on the size
distribution of JTs can be reproduced by a certain set of
collisional parameters. This implies that the collisional
parameters of JT population are different from those of MBA
population.
Another numerical simulation suggests that the impact frequency and relative velocity of objects among MBAs or JTs are not so different \citep{davi02}. 
Therefore, the difference of the shape of size distribution probably caused by a difference of composition and/or internal structure between MBA and JT populations, suggesting they are from different origins.

%There are several points with large error bars, which have less sample namely large uncertainty. 
We also noticed that there is a remarkable dip around a few ten km in the R-plots of Inner and Middle MBAs. The dip becomes shallower in the R-plots of the Outer MBAs, and the dip disappears in the R-plots of the JTs.
%Meanwhile, the dip in the Outer MBAs is shallower than those of Inner/Middle MBAs. For the JTs, there seems no dip around a few ten km in diameter, it is rather flat. By comparing these figures, we can say that the size distribution of JTs is remarkably differ from those of MBAs. 
%For the Outer MBAs, the dip around a few ten km in diameter is quite shallower than those of Inner/Middle MBAs. 
%For the JTs, there is no such dip around a few ten km in diameter. Its R-plot at that size range is rather flat. 
%By comparing these plots, we can say that the size distribution of JTs is remarkably different from those of MBAs. 
%%%%%%%%20170506%%%%%%%%%%%%
%Even among the MBAs, the shape of size distributions are different in each region in the main belt. 
It is well known that a majority of Inner belt objects belong to the S-complex and a majority of Outer belt objects belong to the C-complex. The difference of the size distributions in the main belt regions may be related to the difference of composition and/or inner structure between S- and C-complex. 
%The JT's population mainly consists of D and P-types, which likely have different surface compositions from MBAs. 
% 

\cite{bott05} said that the knee of the size distribution in MBAs around 120 km in diameter is a fossil from the early violent collisional evolution era before the planet migration. 
For the MBAs, the knee around 120 km is shown in all regions: Inner, Middle, and Outer. While, the shape of size distribution of JTs is rather flat around 120 km in diameter.  
This probably suggest that MBAs and JTs have been completely different populations before the planet migration happened.
From these facts mentioned above, we can say that the MBAs and JTs are originated from different regions in their formations.
%}

The shallow dip around a few ten km in the R-plot of the outer main belt may be related to a materials/inner structure difference between the S- and C-complex as mentioned above. 
However, there is another possibility that it may be caused by an influence of implanting objects whose size distribution has no dip around a few ten km such as JTs in the era of LHB, namely the outer objects such as Kuiper Belt objects (KBOs) were captured into the JT region as Nice model suggested, but the objects have even reached till the outer main belt. 
In order to investigate how the outer objects captured into the JT region and the main belt region during the LHB period, further numerical simulations with a high spatial resolution would be needed.

In addition to the simulations, the investigation of the size distribution of small TNOs would be demanded.
%*****
For confirming the prediction from the latest planet migration models, we need a direct comparison of the size distributions between JTs and TNOs. 
However, the surveys of faint TNOs have been insufficient to compare the size distributions with the same size range between JT and TNOs.
\cite{mor09} predicted that the absolute magnitude distribution of the Hot KBOs should become Trojan-like steep at 6$<H$(mag)$<$9 and then \cite{fras14} had reported that the size distribution of JTs is similar with that of the hot KBOs, but their analysis was limited for only large KBOs.
The detection limit of ground-based observation such as Subaru telescope is $r \sim$ 25 mag in usual survey. This means we can detect KBOs larger than $D \sim$110 km for the Classical TNOs assuming at 40 au and albedo of 0.04, or $D \sim$ 60km for the Scattered TNOs assuming at 30 au and albedo of 0.04. 
Thus, it is impossible to find smaller TNOs than 50 km with a capability of the current ground-based telescopes.
Under this situation, the common size range of JTs and TNOs is too narrow to compare properties of the both size distributions. 
Therefore, we need other input source such as the size distribution of craters on icy satellite or Pluto. The New Horizons provided the crater size distribution on Pluto and Charon \citep{sin16}. Their size distributions look flat on the R-plot. The craters would provide us an useful data to derive size distribution of TNOs with much smaller TNOs and then we would be able to compare the size distributions between JTs and TNOs. 
%*****
%

\subsection{Surface number density}
We calculated the surface number density (SND) using the 481 JTs, which is the complete sample set 
from our survey ($H_r$~$\lesssim$~17.4 mag and $r$~$\lesssim$~5.5 au). 
Figure~\ref{fig11} shows the SND of each field with the distance ($d$) of each field from the L4 
point. 
One can see a trend that the SND is larger at the closer field to the L4 point.
The average SNDs for 11.0 $<d$ ($\arcdeg$) $<$ 13.5 (the averaged distance ($\bar{d}$) is 
12.3$\arcdeg$) and for 16.0~$<$~$d$~($\arcdeg$)~$<$~20.0 ($\bar{d}$ is 18.0$\arcdeg$) are 
22.7~$\pm~$3.3~deg$^{-2}$ and 17.9~$\pm$~4.2~deg$^{-2}$, respectively. 
Since our survey of the two fields in $d$~=~14--16$\arcdeg$ had been done with bad seeing, we excluded those fields for the calculation of the SNDs.
Note, our survey area was located at 11--20$\arcdeg$ longitude behind the L4 point. 
While the previous survey by Subaru \,+\, Suprime-Cam done by \cite{YN05, YN08} with almost the same limiting magnitude was done at $\sim$~30$\arcdeg$ in longitude ahead of the L4 point and obtained the SND of 14.8~deg$^{-2}$. 
In order to estimate the population of the entire L4 swarm based on the SNDs, we need further 
observations at different longitudes against the L4 point.

\section{Summary}\label{sec:sum}
We carried out the L4 JT swarm survey by using the Hyper Suprime-Cam attached to the 8.2~m Subaru 
telescope on March 30, 2015 (UT).
We detected 631 JTs in the survey area of $\sim$26~deg$^{2}$ near the opposition and around the 
ecliptic plane with the detection limit of $m_{r}$~=~24.4 mag. 
Our unbiased sample (481 JTs with $H_r$~$\lesssim$~17.4 mag and $r$~$\lesssim$~5.5 au) was used for 
estimating the size distribution of L4 JTs.
Assuming an albedo of 0.07 \citep{grav12}, the size range of the size distribution estimated in this work is corresponding to $D$~$\sim$~2~$-$~20~km.
Our best-fit index ($b$) of cumulative size distribution is $b$~=~1.84~$\pm$~0.05 in $N(>D)~\propto~D^{-b}$.
So far, this work is the deepest survey for L4 JTs determining the size distribution of small L4 JTs with a largest unbiased samples.
%We believe that  we obtained the most robust size distribution of faint L4 JTs because we involved the largest sample. 

Combining L4 JTs detected from our survey with the cataloged L4 JTs, we revealed the size 
distribution of L4 JTs up to $H_r$~=~17.4~mag.

The average surface number density at 12.3$\arcdeg$ and 18.0$\arcdeg$ in longitudes from the L4 
point were found to be 22.7~$\pm$~3.3~deg$^{-2}$ and 17.9~$\pm$~4.2~deg$^{-2}$, respectively.

\section{Acknowledgement}
We are grateful to Keiji Ohtsuki and Naruhisa Takato for very constructive discussions and also thank Takashi Ito, who helped to make figures of size distribution of MBAs with the R-plot (Figure~\ref{fig10} in this paper).
We also thank anonymous referee for providing helpful comments and suggestions.
This publication makes use of data collected by Subaru telescope, which is operated by the National Astronomical Observatory of Japan. 
We also appreciate the Subaru telescope staff and the HSC project staff for their assistance with observations and data reductions.
In this work, we used \textit{hscPipe} (v 3.8.5). Detail information about the software is available at https://arxiv.org/abs/1705.06766.

\clearpage
%%%%%%%%%%%%%%%%%%%%%%%%%%%%%%%%%%%%%%%%%%%%%%%%%%%%%%%%%%%%%%%%%%%%%%%%%%%%%%%%%%%%%%%%%%%%%%%%%%%
%% References
%%%%%%%%%%%%%%%%%%%%%%%%%%%%%%%%%%%%%%%%%%%%%%%%%%%%%%%%%%%%%%%%%%%%%%%%%%%%%%%%%%%%%%%%%%%%%%%%%%%




\clearpage
%%%%%%%%%%%%%%%%%%%%%%%%%%%%%%%%%%%%%%%%%%%%%%%%%%%%%%%%%%%%%%%%%%%%%%%%%%%%%%%%%%%%%%%%%%%%%%%%%%%
%% Table 1
%%%%%%%%%%%%%%%%%%%%%%%%%%%%%%%%%%%%%%%%%%%%%%%%%%%%%%%%%%%%%%%%%%%%%%%%%%%%%%%%%%%%%%%%%%%%%%%%%%%
\begin{deluxetable}{llcccccccc}
\tablecaption{Observation log \label{tab01}}
\tablecolumns{10}
\tablenum{1}
\tablewidth{0pt}
\tablehead{
\colhead{} &
\colhead{} &
\colhead{Time \tablenotemark{a}} & 
\colhead{} &
\colhead{} &
\colhead{} &
\colhead{} &
\colhead{Angle \tablenotemark{b}} &
\colhead{Angle \tablenotemark{c}} & 
\colhead{Number \tablenotemark{d}} \\
\colhead{Field ID} &
\colhead{JD} &
\colhead{interval} & 
\colhead{RA} &
\colhead{Dec} &
\colhead{airmass} &
\colhead{seeing} &
\colhead{from} &
\colhead{from} & 
\colhead{of} \\
\colhead{} & 
\colhead{(day)} &
\colhead{(min)} & 
\colhead{(deg)} &
\colhead{(deg)} &
\colhead{} & 
\colhead{(arcsec)} & 
\colhead{opp. (deg)} &
\colhead{L4 (deg)} &
\colhead{detection}
}
\startdata
FIELD01	&	2457111.823148 	&	18.250 	&	184.14960 	&	-1.69997 	&	1.335 	&	0.539 	&	4.8    	&	19.8  	&	33	\\
		&	2457111.835822 	&	55.850 	&	184.14960 	&	-1.69997 	&	1.265 	&	0.590 	&			&			&		\\
		&	2457111.874606 	&			&	184.14958 	&	-1.69999 	&	1.129 	&	0.666 	&			&			&		\\
FIELD02	&	2457111.826343 	&	18.200 	&	184.89958 	&	-3.00000 	&	1.343 	&	0.558 	&	3.7 	        &	18.6  	&	33	\\
		&	2457111.838981 	&	55.900 	&	184.89958 	&	-3.00001 	&	1.273 	&	0.971 	&			&			&		\\
		&	2457111.877801 	&			&	184.89961 	&	-2.99997 	&	1.137 	&	0.690 	&			&			&		\\
FIELD03	&	2457111.829502 	&	18.267 	&	185.64960 	&	-1.69998 	&	1.322 	&	0.551 	&	3.5 	        &	18.3  	&	39	\\
		&	2457111.842188 	&	55.933 	&	185.64960 	&	-1.69996 	&	1.254 	&	0.882 	&			&			&		\\
		&	2457111.881030 	&			&	185.64960 	&	-1.69997 	&	1.124 	&	0.589 	&			&			&		\\
FIELD04	&	2457111.832662 	&	18.350 	&	186.39959 	&	-2.99999 	&	1.330 	&	0.553 	&	2.2    	&	17.2  	&	35	\\
		&	2457111.845405 	&	55.950 	&	186.39960 	&	-2.99999 	&	1.262 	&	0.836 	&			&			&		\\
		&	2457111.884259 	&			&	186.39960 	&	-3.00000 	&	1.132 	&	0.586 	&			&			&		\\
FIELD05	&	2457111.848600 	&	18.350 	&	187.14961 	&	-1.69995 	&	1.243 	&	1.092 	&	2.4    	&	16.9  	&	31	\\
		&	2457111.861343 	&	37.667 	&	187.14960 	&	-1.69999 	&	1.191 	&	0.724 	&			&			&		\\
		&	2457111.887500 	&			&	187.14962 	&	-1.69998 	&	1.119 	&	0.621 	&			&			&		\\
FIELD06	&	2457111.851794 	&	18.883 	&	186.39958 	&	-0.39999 	&	1.208 	&	0.761 	&	3.9    	&	18.2  	&	28	\\
		&	2457111.864907 	&	37.067 	&	186.39961 	&	-0.39998 	&	1.161 	&	0.788 	&			&			&		\\
		&	2457111.890648 	&			&	186.39959 	&	-0.39997 	&	1.099 	&	0.549 	&			&			&		\\
FIELD07	&	2457111.854988 	&	18.867 	&	187.89958 	&	-0.39999 	&	1.212 	&	0.775 	&	3.3    	&	16.8  	&	48	\\
		&	2457111.868090 	&	37.083 	&	187.89961 	&	-0.39997 	&	1.164 	&	0.795 	&			&			&		\\
		&	2457111.893843 	&			&	187.89961 	&	-0.39996 	&	1.101 	&	0.566 	&			&			&		\\
FIELD08	&	2457111.858148 	&	18.883 	&	189.39960 	&	-0.39998 	&	1.217 	&	0.839 	&	3.4    	&	15.5  	&	44	\\
		&	2457111.871262 	&	37.067 	&	189.39960 	&	-0.39999 	&	1.168 	&	0.685 	&			&			&		\\
		&	2457111.897002 	&			&	189.39961 	&	-0.39998 	&	1.103 	&	0.566 	&			&			&		\\
FIELD09	&	2457111.900949 	&	18.367 	&	187.89962 	&	-2.99998 	&	1.110 	&	0.531 	&	0.9    	&	15.8  	&	43	\\
		&	2457111.913704 	&	50.600 	&	187.89992 	&	-2.99969 	&	1.094 	&	0.624 	&			&			&		\\
		&	2457111.948843 	&		 	&	187.90060 	&	-3.00004 	&	1.091 	&	0.898 	&			&			&		\\
FIELD10	&	2457112.047199 	&	18.550 	&	188.64960 	&	-1.69998 	&	1.415 	&	1.173 	&	2.0    	&	15.6  	&	23	\\
		&	2457112.060081 	&	55.150 	&	188.64960 	&	-1.70000 	&	1.526 	&	1.405 	&			&			&		\\
		&	2457112.098380 	&			&	188.64958 	&	-1.69996 	&	2.095 	&	1.033 	&			&			&		\\
FIELD11	&	2457112.050417 	&	18.517 	&	189.39961 	&	-3.00001 	&	1.441 	&	1.239 	&	1.1    	&	14.4  	&	14	\\
		&	2457112.063275 	&	55.117 	&	189.39961 	&	-3.00001 	&	1.557 	&	1.298 	&			&			&		\\
		&	2457112.101551 	&			&	189.39963 	&	-3.00001 	&	2.157 	&	1.746 	&			&			&		\\
FIELD12	&	2457112.053634 	&	18.467 	&	190.89961 	&	-3.00000 	&	1.433 	&	1.098 	&	2.5 	        &	13.0  	&	33	\\
		&	2457112.066458 	&	55.133 	&	190.89959 	&	-2.99998 	&	1.547 	&	0.948 	&			&			&		\\
		&	2457112.104745 	&			&	190.89959 	&	-2.99997 	&	2.134 	&	1.433 	&			&			&		\\
FIELD13	&	2457112.056840 	&	18.417 	&	191.64958 	&	-1.69997 	&	1.425 	&	1.402 	&	3.7    	&	12.9  	&	33	\\
		&	2457112.069630 	&	55.150 	&	191.64958 	&	-1.69995 	&	1.538 	&	0.935 	&			&			&		\\
		&	2457112.107928 	&			&	191.64959 	&	-1.70001 	&	2.123 	&	1.156 	&			&			&		\\
FIELD14	&	2457112.072824 	&	18.433 	&	192.39961 	&	-2.99998 	&	1.570 	&	0.865 	&	3.9    	&	11.6          &	44	\\
		&	2457112.085625 	&	36.667 	&	192.39957 	&	-2.99994 	&	1.722 	&	0.849 	&			&			&		\\
		&	2457112.111088 	&			&	192.39960 	&	-2.99999 	&	2.186 	&	1.335 	&			&			&		\\
FIELD15	&	2457112.076053 	&	18.400 	&	191.64959 	&	-4.29998 	&	1.651 	&	0.917 	&	3.2 	        &	11.8  		&	52	\\
		&	2457112.088831 	&	36.617 	&	191.64961 	&	-4.30002 	&	1.826 	&	1.051 	&			&			&		\\
		&	2457112.114259 	&			&	191.64961 	&	-4.30003 	&	2.370 	&	1.159 	&			&			&		\\
FIELD16	&	2457112.079248 	&	18.350 	&	191.19961 	&	-5.75000 	&	1.736 	&	0.978 	&	3.4     	&	11.9  		&	48	\\
		&	2457112.091991 	&	36.633 	&	191.19961 	&	-5.75001 	&	1.934 	&	1.058 	&			&			&		\\
		&	2457112.117431 	&			&	191.19958 	&	-5.74996 	&	2.568 	&	1.228 	&			&			&		\\
FIELD17	&	2457112.082477 	&	18.250 	&	190.79960 	&	-7.20002 	&	1.831 	&	0.910 	&	4.2    	&	12.0  	&	50	\\
		&	2457112.095150 	&	36.650 	&	190.79961 	&	-7.20001 	&	2.055 	&	1.016 	&			&			&		\\
		&	2457112.120602 	&			&	190.79959 	&	-7.19999 	&	2.801 	&	1.159 	&			&			&		\\
\enddata
\tablenotetext{a}{Time interval between first and second visits and between second and third visits.}
\tablenotetext{b}{Average angle between the center of position for each field and the opposition ($\lambda$, $\beta$)=(189.3, 0.0).}
\tablenotetext{c}{Average angle between the center of position for each field and the L4 point ($\lambda$, $\beta$)=(204.2, 1.6)}
\tablenotetext{d}{Number of detected JTs with the three visits.}
\end{deluxetable}

\clearpage
%%%%%%%%%%%%%%%%%%%%%%%%%%%%%%%%%%%%%%%%%%%%%%%%%%%%%%%%%%%%%%%%%%%%%%%%%%%%%%%%%%%%%%%%%%%%%%%%%%%
%% Figure 1
%%%%%%%%%%%%%%%%%%%%%%%%%%%%%%%%%%%%%%%%%%%%%%%%%%%%%%%%%%%%%%%%%%%%%%%%%%%%%%%%%%%%%%%%%%%%%%%%%%%
\begin{figure}
\figurenum{1}
\plotone{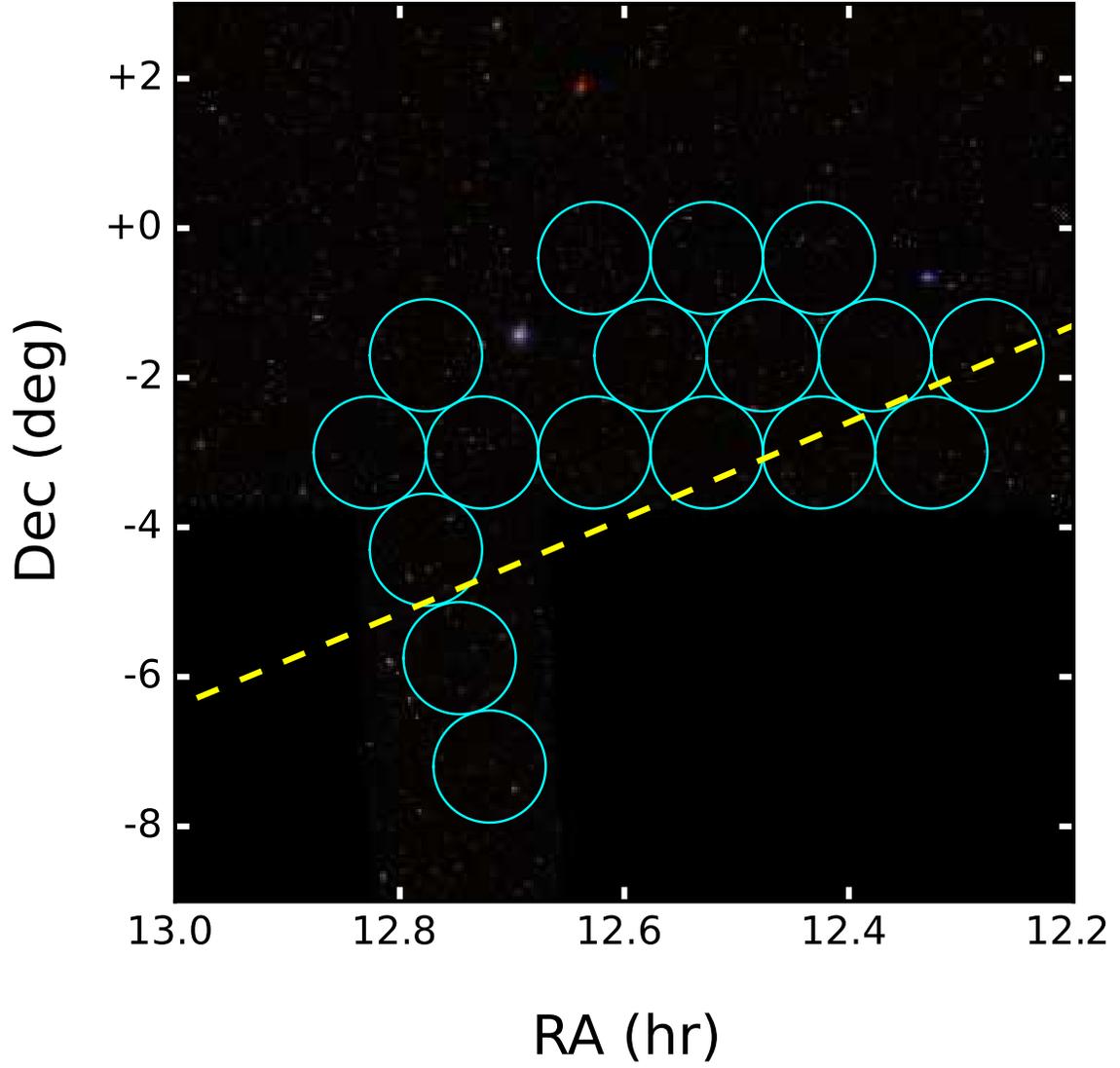}
\caption{
The observation area of our survey. 
Every circle corresponds to each of HSC's field-of-view (FOV). 
The background image is from SDSS-DR9 \citep{alam15}. 
The broken line shows the ecliptic plane. The L4 point's location is outside the image.
\label{fig01}
}
\end{figure}

\clearpage
%%%%%%%%%%%%%%%%%%%%%%%%%%%%%%%%%%%%%%%%%%%%%%%%%%%%%%%%%%%%%%%%%%%%%%%%%%%%%%%%%%%%%%%%%%%%%%%%%%%
%% Figure 2
%%%%%%%%%%%%%%%%%%%%%%%%%%%%%%%%%%%%%%%%%%%%%%%%%%%%%%%%%%%%%%%%%%%%%%%%%%%%%%%%%%%%%%%%%%%%%%%%%%%
\begin{figure}
\figurenum{2}
\plotone{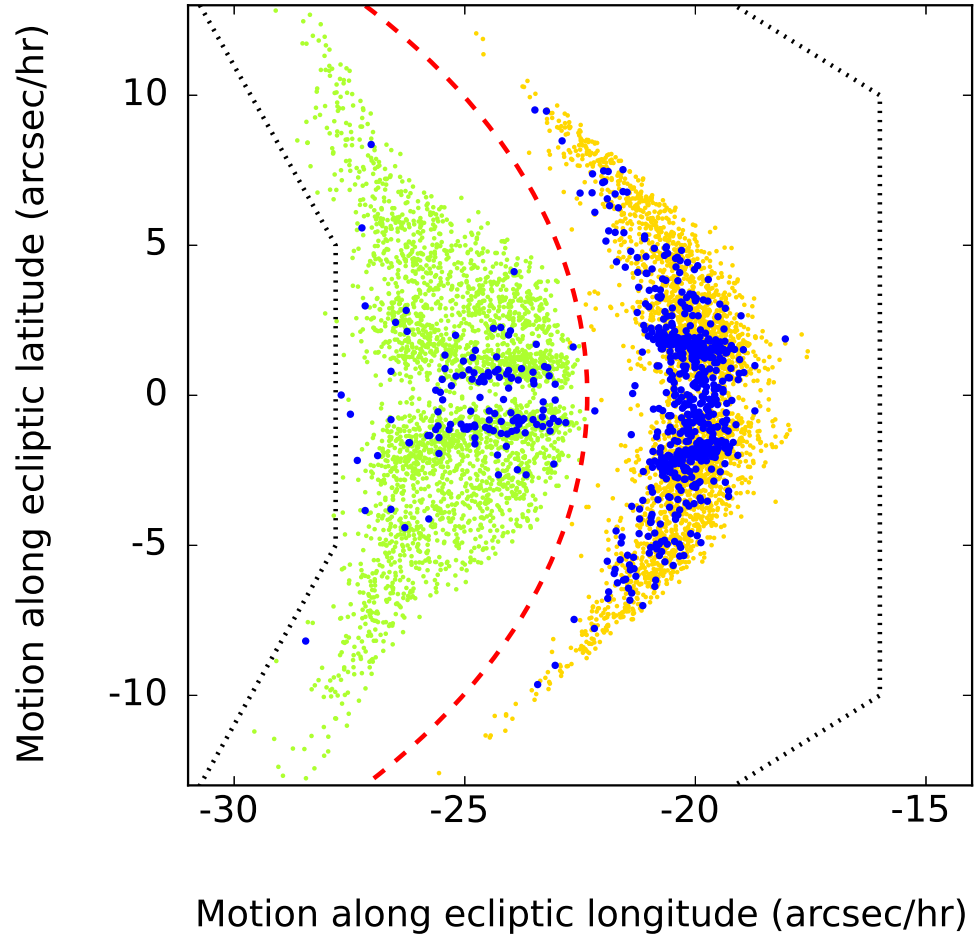}
\caption{
Motions along ecliptic longitude/latitude of detected moving objects (blue dots). 
The lime and orange dots show those of artificial Hildas and JTs, respectively, generated based on 
their orbital distributions of known objects by Monte-Carlo method. 
The area surrounded with dotted lines (black) is the motion range of our search.
The broken line (red) shows the boundary between Hildas and JTs. 
One can see that the moving objects detected from our survey were clearly divided into two groups : Hildas and JTs. We could pick up JTs without contamination of Hildas.
\label{fig02}
}
\end{figure}

\clearpage
%%%%%%%%%%%%%%%%%%%%%%%%%%%%%%%%%%%%%%%%%%%%%%%%%%%%%%%%%%%%%%%%%%%%%%%%%%%%%%%%%%%%%%%%%%%%%%%%%%%
%% Figure 3
%%%%%%%%%%%%%%%%%%%%%%%%%%%%%%%%%%%%%%%%%%%%%%%%%%%%%%%%%%%%%%%%%%%%%%%%%%%%%%%%%%%%%%%%%%%%%%%%%%%
\begin{figure}
\figurenum{3}
\plotone{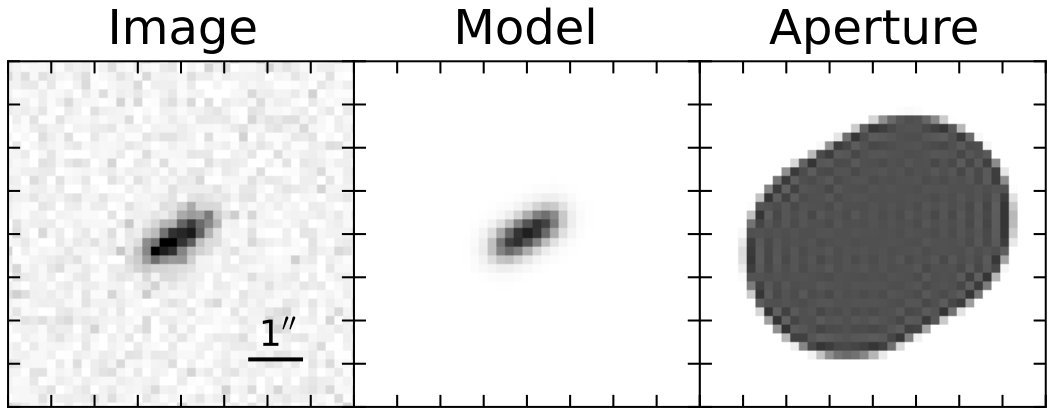}
\caption{
An image of a detected JT (left), the best-fit model (middle), and the weight 
coefficient map for aperture photometry based on the sinc integration technique (right). 
All the window have the same size of 7$\arcsec$~$\times$~7$\arcsec$. 
\label{fig03}
}
\end{figure}

\clearpage
%%%%%%%%%%%%%%%%%%%%%%%%%%%%%%%%%%%%%%%%%%%%%%%%%%%%%%%%%%%%%%%%%%%%%%%%%%%%%%%%%%%%%%%%%%%%%%%%%%%
%% Figure 4
%%%%%%%%%%%%%%%%%%%%%%%%%%%%%%%%%%%%%%%%%%%%%%%%%%%%%%%%%%%%%%%%%%%%%%%%%%%%%%%%%%%%%%%%%%%%%%%%%%%
\begin{figure}
\figurenum{4}
\plotone{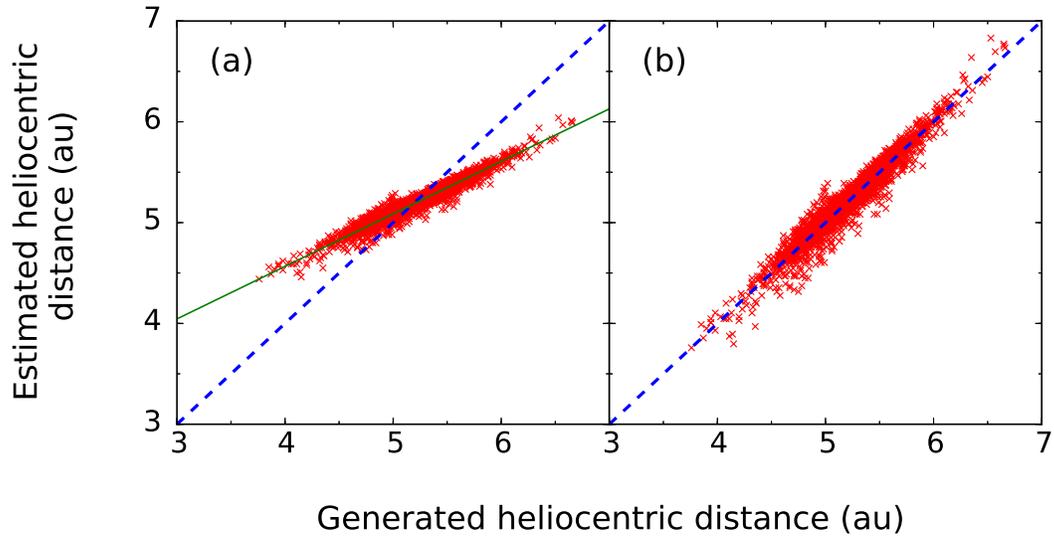}
\caption{
(a) Correlation plot between the generated and estimated heliocentric distances (see text). 
The solid line shows the best-fit linear function.
(b) The same plot as (a), but the ordinate represents the corrected heliocentric distance. 
\label{fig04}
}
\end{figure}

\clearpage
%%%%%%%%%%%%%%%%%%%%%%%%%%%%%%%%%%%%%%%%%%%%%%%%%%%%%%%%%%%%%%%%%%%%%%%%%%%%%%%%%%%%%%%%%%%%%%%%%%%
%% Figure 5
%%%%%%%%%%%%%%%%%%%%%%%%%%%%%%%%%%%%%%%%%%%%%%%%%%%%%%%%%%%%%%%%%%%%%%%%%%%%%%%%%%%%%%%%%%%%%%%%%%%
\begin{figure}
\figurenum{5}
\plotone{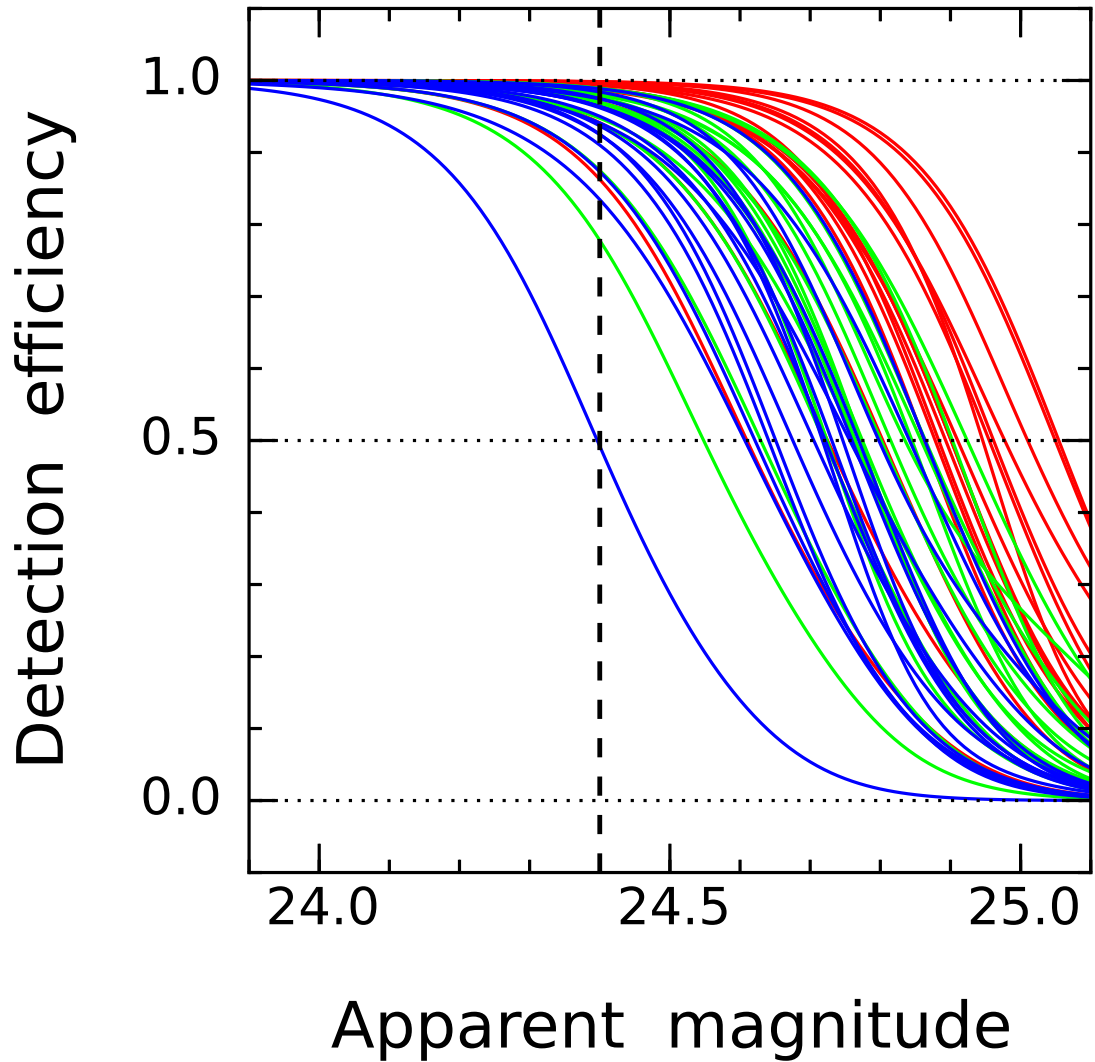}
\caption{
The detection efficiency of each observational field. Red, green, blue lines stand for the 
detection efficiency on the CCD near the center, middle and edge of each field, respectively. 
Each color contains 17 lines corresponding to 17 observational FIELDs.
We defined the detection limit with the lowest detection efficiency of 0.5 (i.e. 50 \% detection), 
which is 24.4 mag in apparent magnitude (vertical broken line).
\label{fig05}
}
\end{figure}

\clearpage
%%%%%%%%%%%%%%%%%%%%%%%%%%%%%%%%%%%%%%%%%%%%%%%%%%%%%%%%%%%%%%%%%%%%%%%%%%%%%%%%%%%%%%%%%%%%%%%%%%%
%% Figure 6
%%%%%%%%%%%%%%%%%%%%%%%%%%%%%%%%%%%%%%%%%%%%%%%%%%%%%%%%%%%%%%%%%%%%%%%%%%%%%%%%%%%%%%%%%%%%%%%%%%%
\begin{figure}
\figurenum{6}
\plotone{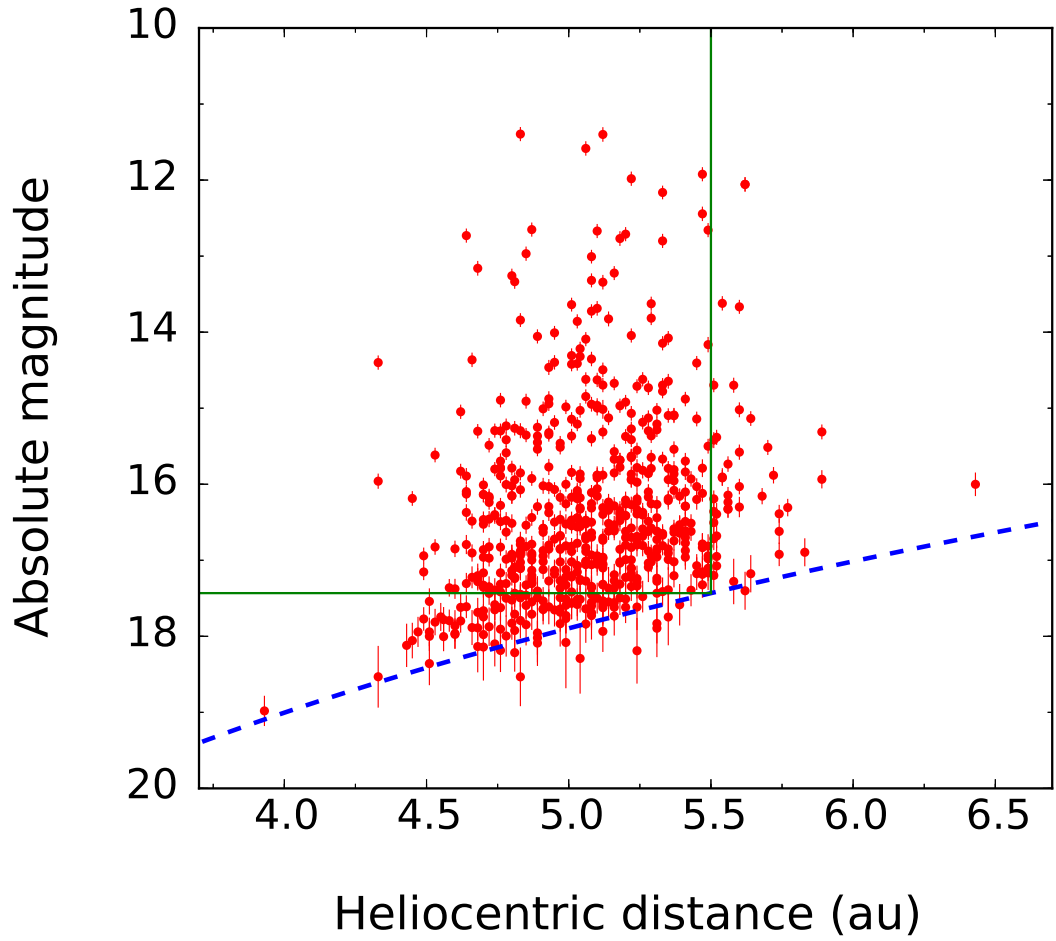}
\caption{
Heliocentric distance ($r$) vs. absolute magnitude ($H$) for the detected JTs.
The broken line corresponds to apparent magnitude of 24.4~mag. 
The solid lines show the border of unbiased sample, $r$~=~5.5~au and $H$~=~17.4~mag.  The data points shown in this figure is available as the Data behind the Figure.
\label{fig06}
}
\end{figure}

\clearpage
%%%%%%%%%%%%%%%%%%%%%%%%%%%%%%%%%%%%%%%%%%%%%%%%%%%%%%%%%%%%%%%%%%%%%%%%%%%%%%%%%%%%%%%%%%%%%%%%%%%
%% Figure 7
%%%%%%%%%%%%%%%%%%%%%%%%%%%%%%%%%%%%%%%%%%%%%%%%%%%%%%%%%%%%%%%%%%%%%%%%%%%%%%%%%%%%%%%%%%%%%%%%%%%
\begin{figure}
\figurenum{7}
\plotone{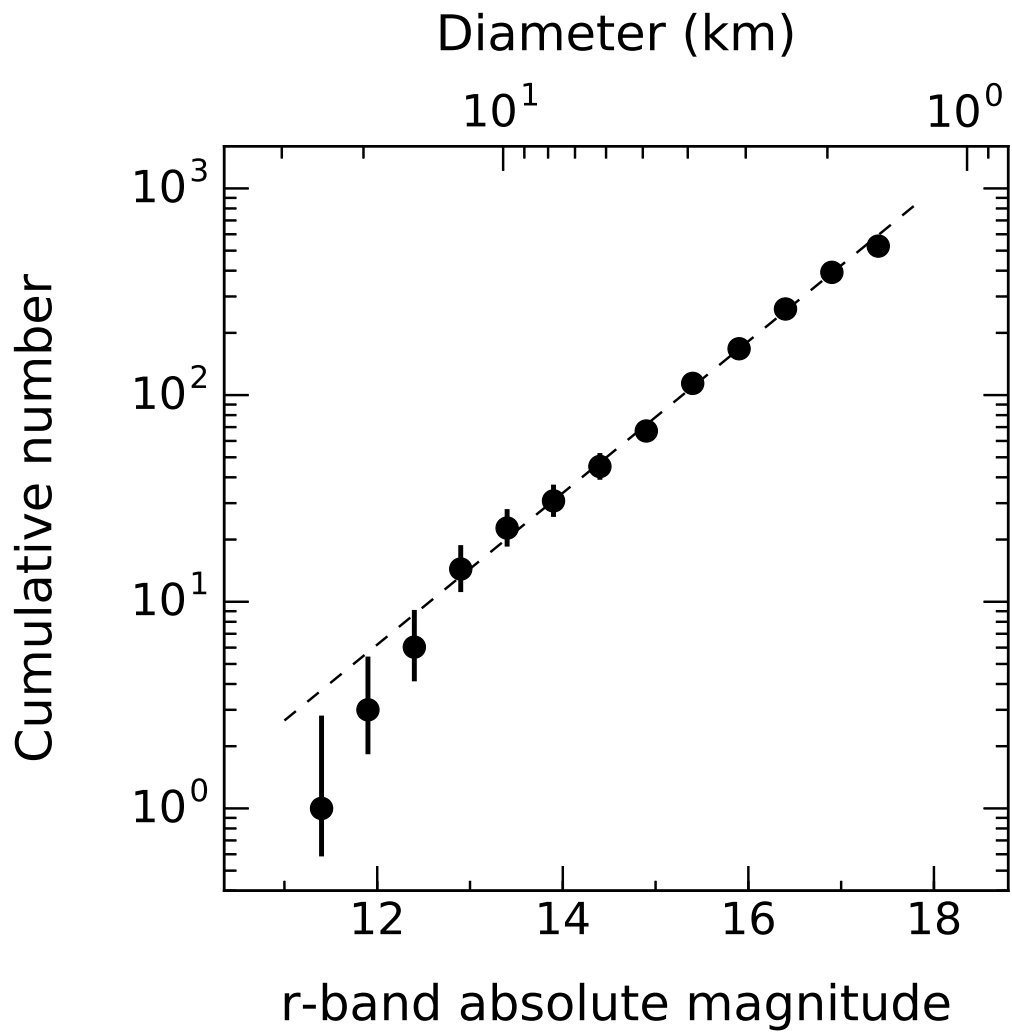}
\caption{
The cumulative size distribution for the 431 JTs obtained from this survey. 
The broken line shows the best-fit power law approximation with an index of $\alpha$~=~0.37 or $b$~=~1.84 
(see text). 
The data points shown in this figure is available as the Data behind the Figure.
\label{fig07}
}
\end{figure}

\clearpage
%%%%%%%%%%%%%%%%%%%%%%%%%%%%%%%%%%%%%%%%%%%%%%%%%%%%%%%%%%%%%%%%%%%%%%%%%%%%%%%%%%%%%%%%%%%%%%%%%%%
%% Figure 8
%%%%%%%%%%%%%%%%%%%%%%%%%%%%%%%%%%%%%%%%%%%%%%%%%%%%%%%%%%%%%%%%%%%%%%%%%%%%%%%%%%%%%%%%%%%%%%%%%%%
\begin{figure}
\figurenum{8}
\plotone{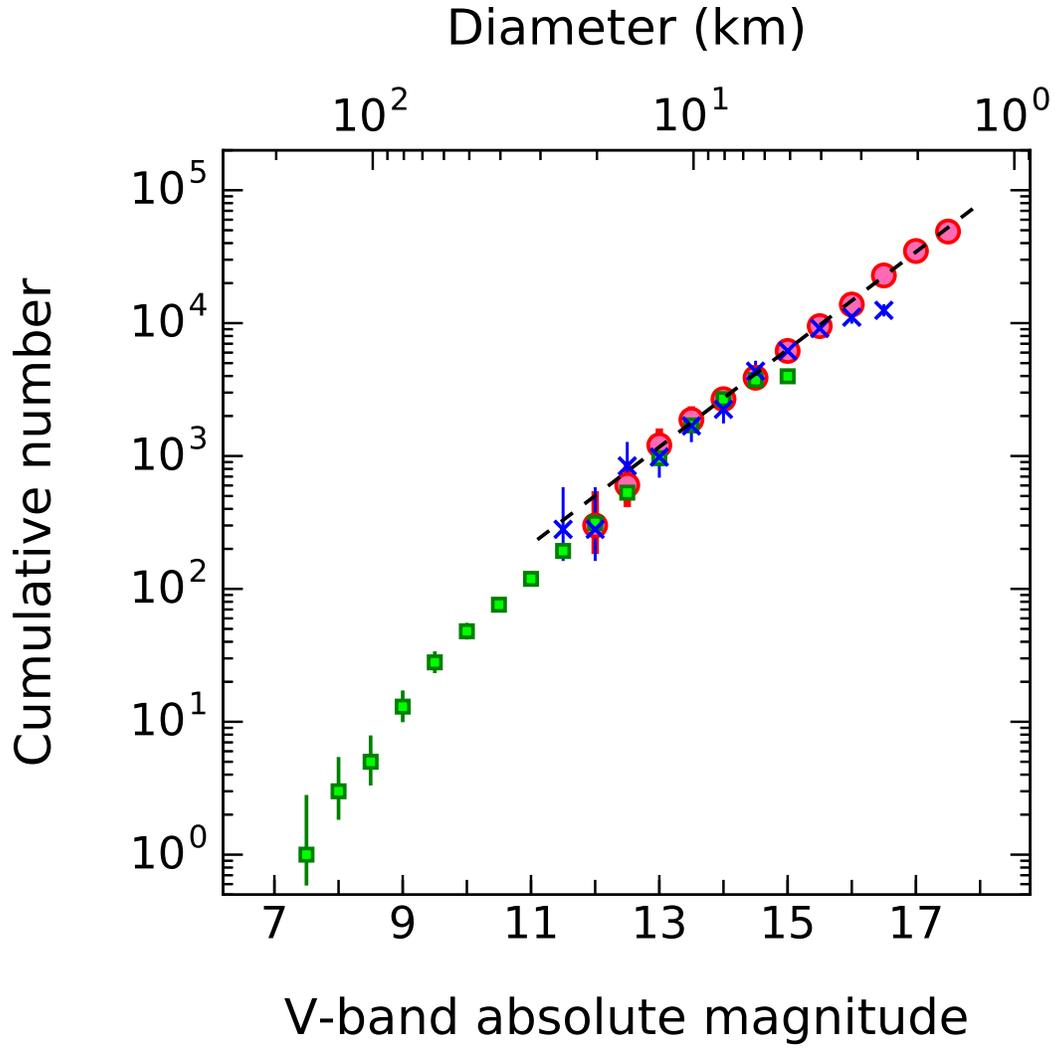}
\caption{
The cumulative size distributions of L4 JTs combined with the MPC catalog (squares) and this work
(circles) scaled at $H_V$~=~14.0~mag.
The $V-r$ color is assumed to be 0.25~mag \citep{szabo07}.
The size distribution data from \citet{jew00} scaled at $H_V$~=~15.0~mag is also plotted 
(crosses).
The broken line shows the same power law approximation as Figure~\ref{fig07}.
\label{fig08}
}
\end{figure}

\clearpage
%%%%%%%%%%%%%%%%%%%%%%%%%%%%%%%%%%%%%%%%%%%%%%%%%%%%%%%%%%%%%%%%%%%%%%%%%%%%%%%%%%%%%%%%%%%%%%%%%%%
%% Figure 9
%%%%%%%%%%%%%%%%%%%%%%%%%%%%%%%%%%%%%%%%%%%%%%%%%%%%%%%%%%%%%%%%%%%%%%%%%%%%%%%%%%%%%%%%%%%%%%%%%%%
\begin{figure}
\figurenum{9}
\plotone{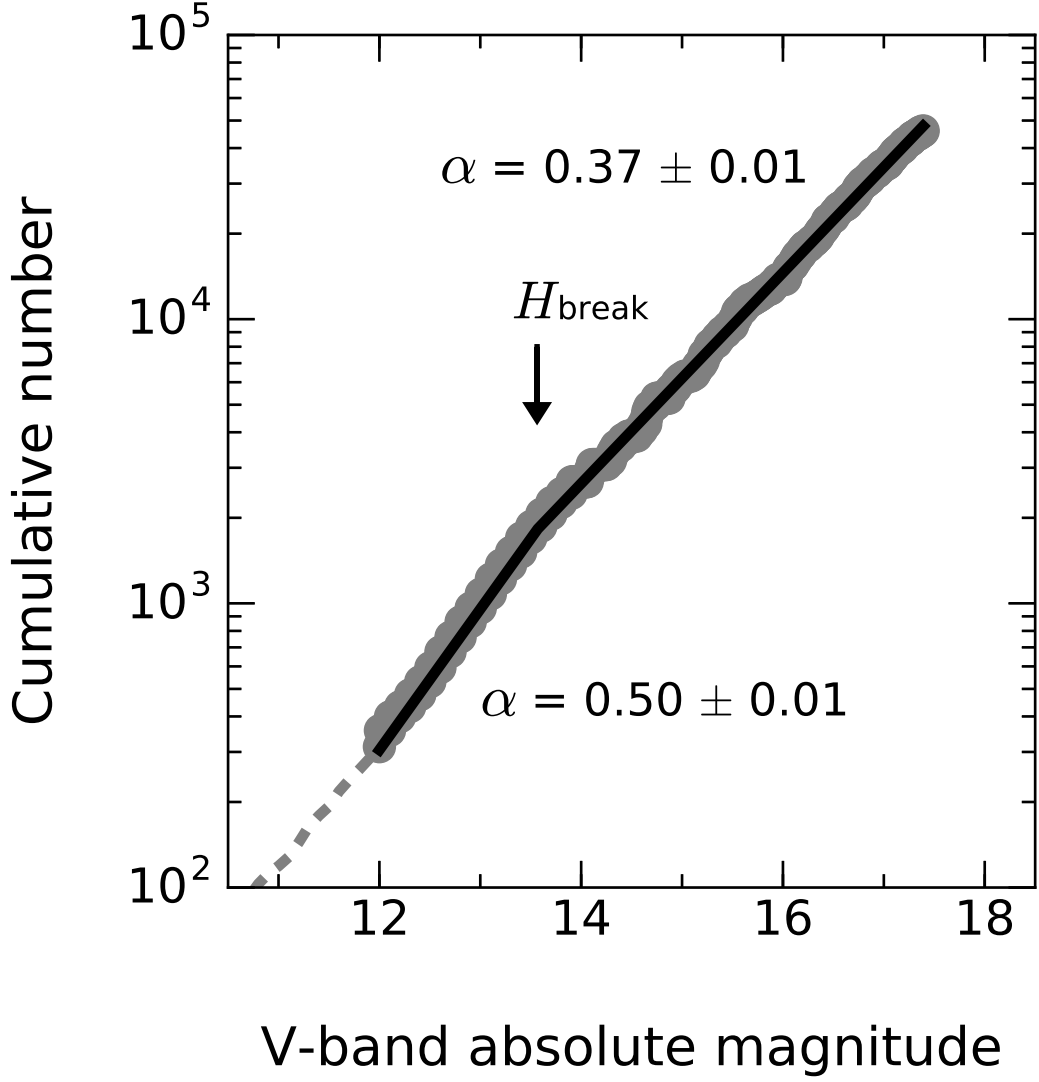}
\caption{
%\textbf{
%The best-fit broken power law (solid line) to the size distribution combined with the MPC catalog and this work scaled at $H_V$~=~14.0~mag (gray dots) in $H_V$~$>$~12.0~mag.
A broken power law function that fits the best to the cumulative size distribution which is a combination of the MPC catalog and this work scaled at $H_V$~=~14.0~mag (gray dots) in $H_V$~$>$~12.0~mag.
The power-law indexes are $\alpha$~=~0.50~$\pm$~0.01 for the brighter objects and
$\alpha$~=~0.37~$\pm$~0.01 for the fainter objects.
The break point is located at $H_{\rm break}$~=~13.56$^{+0.04}_{-0.06}$~mag.
%}
\label{fig09}
}
\end{figure}

\clearpage
%%%%%%%%%%%%%%%%%%%%%%%%%%%%%%%%%%%%%%%%%%%%%%%%%%%%%%%%%%%%%%%%%%%%%%%%%%%%%%%%%%%%%%%%%%%%%%%%%%%
%% Figure 10
%%%%%%%%%%%%%%%%%%%%%%%%%%%%%%%%%%%%%%%%%%%%%%%%%%%%%%%%%%%%%%%%%%%%%%%%%%%%%%%%%%%%%%%%%%%%%%%%%%%
\begin{figure}
\figurenum{10}
\plotone{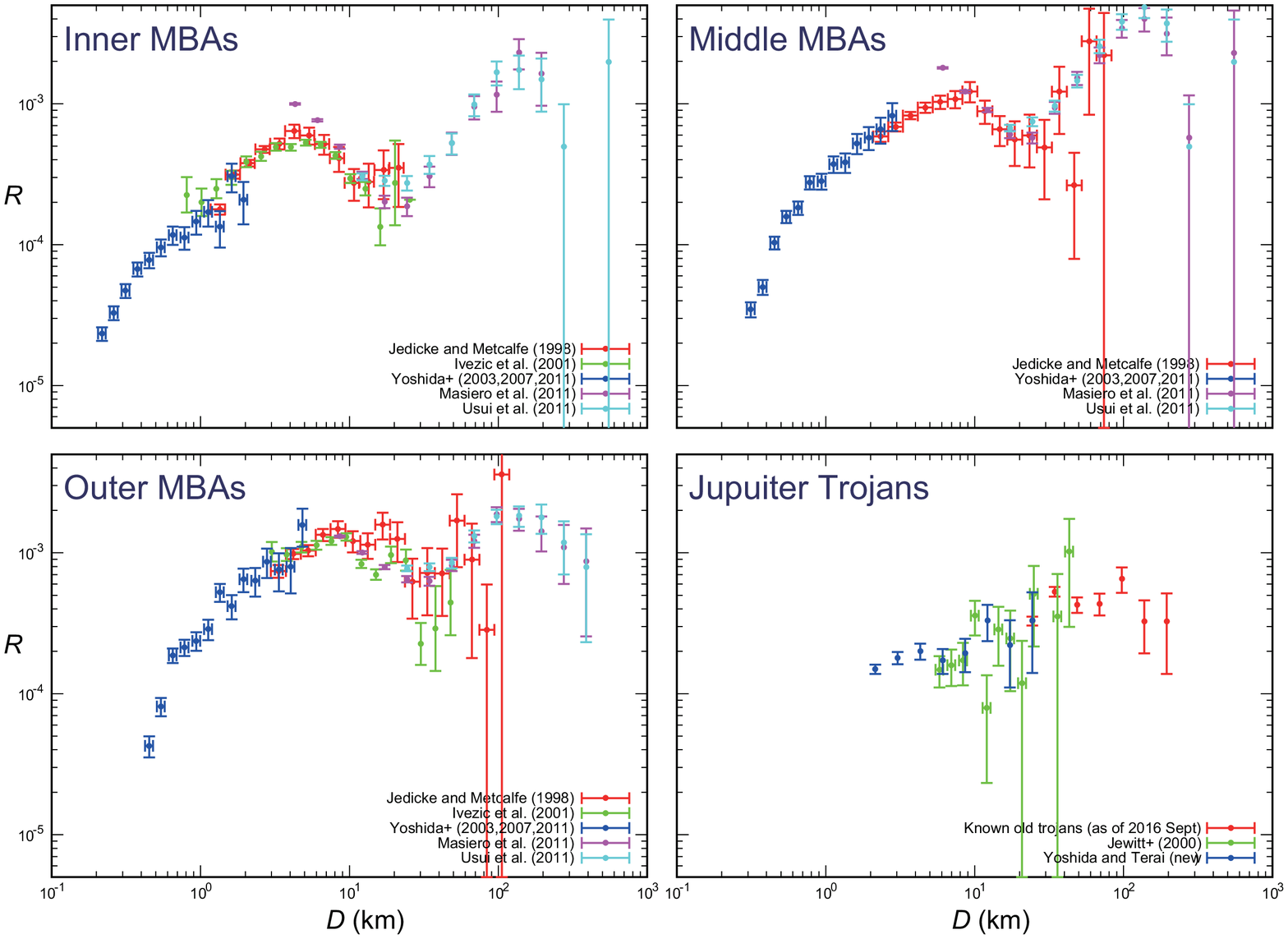}
\caption{
The size distributions of MBAs and JT with the R-plot. The data for MBAs were collected by Spacewatch survey (red)  \citep{JM98}, Sloan Dedital Sky Survey (SDSS) (green) \citep{ivez01}, Subaru telescope (blue)  \citep{yoshi03, YN07, yoshi11}, WISE (magenta)  \citep{masi11} and AKARI (sky blue) \citep{usui11}. The data for JTs were collected by our survey with Subaru telescope (blue), the survey done by \citet{jew00} (green), and known JTs with H$<$12.3 mag (red) 
which have been discovered and listed in the ASTORB catalog as suggested by \citet{szabo07}. 
We divided the main belt region into the three parts : Inner 2.0~$<$~$a$ (au)~$<$~2.6, Middle 2.6~$<$~$a$ (au)~$<$~3.0, Outer 3.0~$<$~$a$ (au)~$<$~3.5. 
For the data sets from the Spacewatch survey and Subaru telescope (SMBAS), we assumed albedo 0.16, 0.13, and 0.10 for Inner, Middle, and Outer MBAs, respectively. These albedo were calculated using the average albedo of S-complex (0.21) and C-complex (0.05) and the S-complex/C-complex ratio in each main belt region obtained by \cite{YN07}, which have been used in \cite{strom15}. 
As for SDSS data set, \cite{ivez01} provided the diameter distribution of only red (S-complex) and blue (C-complex) groups in the entire main belt. Therefore we included the size distribution of the red group into the plot of the Inner MBAs and that of the blue group into the plot of the outer MBAs. In the plot of hte Middle MBAs, there is no data from SDSS. 
We considered the detection limit from each survey and determined the range of the plot.
\label{fig10}
}
\end{figure}

\clearpage
%%%%%%%%%%%%%%%%%%%%%%%%%%%%%%%%%%%%%%%%%%%%%%%%%%%%%%%%%%%%%%%%%%%%%%%%%%%%%%%%%%%%%%%%%%%%%%%%%%%
%% Figure 11
%%%%%%%%%%%%%%%%%%%%%%%%%%%%%%%%%%%%%%%%%%%%%%%%%%%%%%%%%%%%%%%%%%%%%%%%%%%%%%%%%%%%%%%%%%%%%%%%%%%
\begin{figure}
\figurenum{11}
\plotone{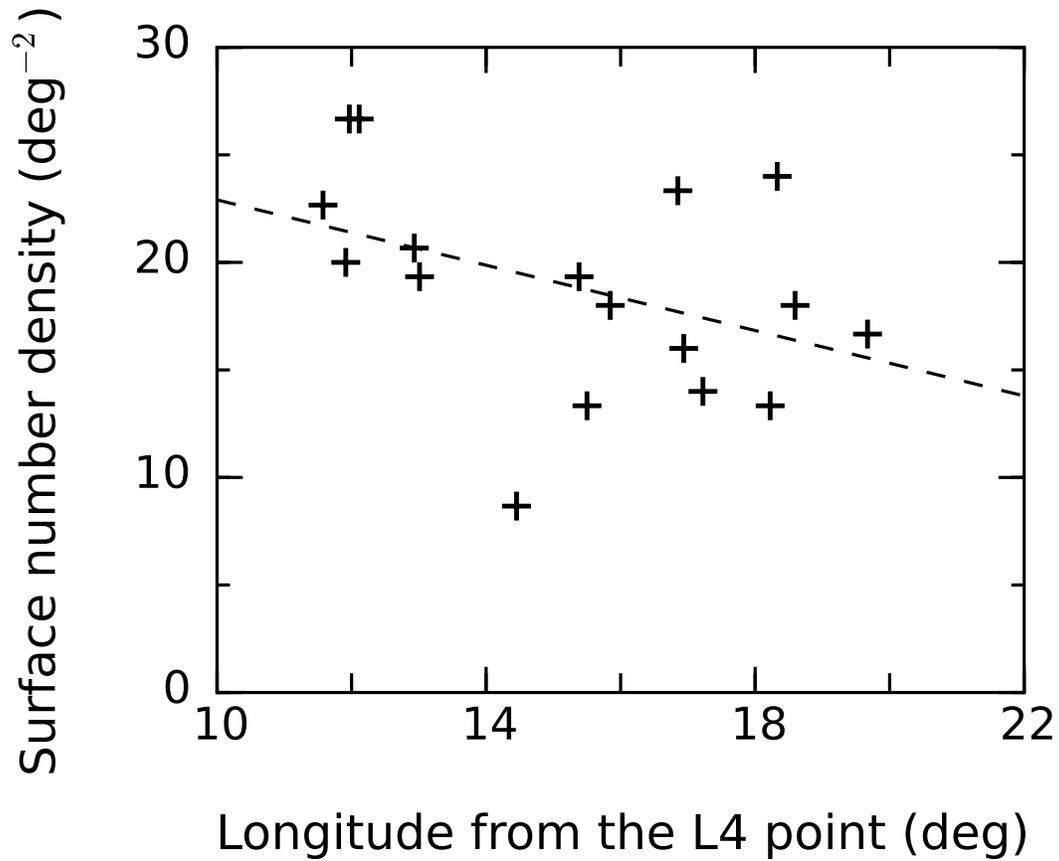}
\caption{
The surface number density (SND) with longitude from the L4 point. $+$ stands for the SND of each field. 
As shown in the orbital distribution of known JTs, the SND can be larger as closer to the L4 point. 
Since each field has different observational conditions (e.g. different seeing size, transparency), the distribution of SNDs in our survey are a little scattered. However, we can see a trend (broken line) that the SND is larger as closer to L4 point.
\label{fig11}
}
\end{figure}



\end{document}